\documentclass[%
 reprint,
superscriptaddress,
 amsmath,amssymb,
 aps,
]{revtex4-2}

\usepackage{graphicx}% Include figure files
\usepackage{dcolumn}% Align table columns on decimal point
\usepackage{bm}% bold math
\usepackage{mathtools}
\DeclarePairedDelimiterX\braket[2]{\langle}{\rangle}{#1\,\delimsize\vert\,\mathopen{}#2}

\usepackage{color}

\begin{document}

\preprint{APS/123-QED}

\title{Quasinormal coupled-mode analysis of dynamic gain in exceptional-point lasers}% Force line breaks with \\

\author{Hao He}
\email{hehao@usc.edu}
\author{Xingwei Gao}%
\affiliation{%
 Ming Hsieh Department of Electrical and Computer Engineering, University of Southern California, Los Angeles, CA 90007, USA}%
 
\author{Alexander Cerjan}
\email{awcerja@sandia.gov}
\affiliation{
 Center for Integrated Nanotechnologies, Sandia National Laboratories, Albuquerque, NM 87185, USA
}%
\author{Chia Wei Hsu}
\affiliation{%
 Ming Hsieh Department of Electrical and Computer Engineering, University of Southern California, Los Angeles, CA 90007, USA}%

\

\begin{abstract}
One of the key features of lasers operating near exceptional points (EPs) is that the gain medium can support an oscillating population inversion above a pump threshold, leading to self-modulated laser dynamics.
This unusual behavior opens up new possibilities for frequency comb generation and temporal modulation. 
However, the dynamic population inversion couples signals with different frequencies and thus cannot be captured by conventional temporal coupled-mode theory (TCMT) based on static saturable gain. 
In this paper, we develop a perturbative coupled-mode analysis framework to capture the spatial-temporal dynamics of near-EP lasers. 
By decomposing discrete frequency generation into multiple excitations of resonant modes, our analysis establishes a minimal physical model that translates the local distribution of dynamic population-inversion into a resonant modal interpretation of laser gain.
This work enables the exploration of unique properties in this self-time-modulated systems, such as time-varying scattering and non-reciprocal transmission.
\end{abstract}

\maketitle

%\tableofcontents

\section{\label{sec:introduction}Introduction}
Non-Hermiticity is a crucial feature of optical systems due to the ubiquity of both material absorption and radiative losses to the surrounding environment \cite{el2018non}.
Among various non-Hermitian effects, exceptional points (EP)---spectral singularities in non-Hermitian systems---are crucial to understand because they fundamentally alter a system's response and promote a variety of optical applications, such as enhanced sensitivity and chiral control of light
\cite{miri2019exceptional,ozdemir2019parity,li2023exceptional,hodaei2017enhanced,hokmabadi2019non,lai2019observation,wiersig2008asymmetric,wiersig2011nonorthogonal,peng2016chiral,zhang2020tunable}.
Lasers have been an ideal platform to demonstrate key characteristics of EPs as they exhibit a natural way to implement spatially non-uniform gain and loss, as is required for EP formation \cite{zhao2018parity,ruter2010observation,peng2014parity,liertzer2012pump,cerjan2016eigenvalue}.
Recent studies found that the interplay between EPs and lasers reveals exciting opportunities to control laser dynamics, such as loss-induced lasing \cite{peng2014loss}, reverse pump dependence \cite{liertzer2012pump,brandstetter2014reversing}, and robust single-mode operation \cite{feng2014single,hodaei2014parity}.
Generally, only one mode among the paired resonances of an EP lases at a single time and so the carrier populations of the gain medium can be approximated as stationary.
In these cases, the other latent resonances(s) of an EP can lead to undesired laser linewidth broadening while maintaining single mode operation \cite{zhang2018phonon,wang2020petermann,benzaouia2022nonlinear}.

\begin{figure*}[t]
\centering
\includegraphics[width=0.8\textwidth]{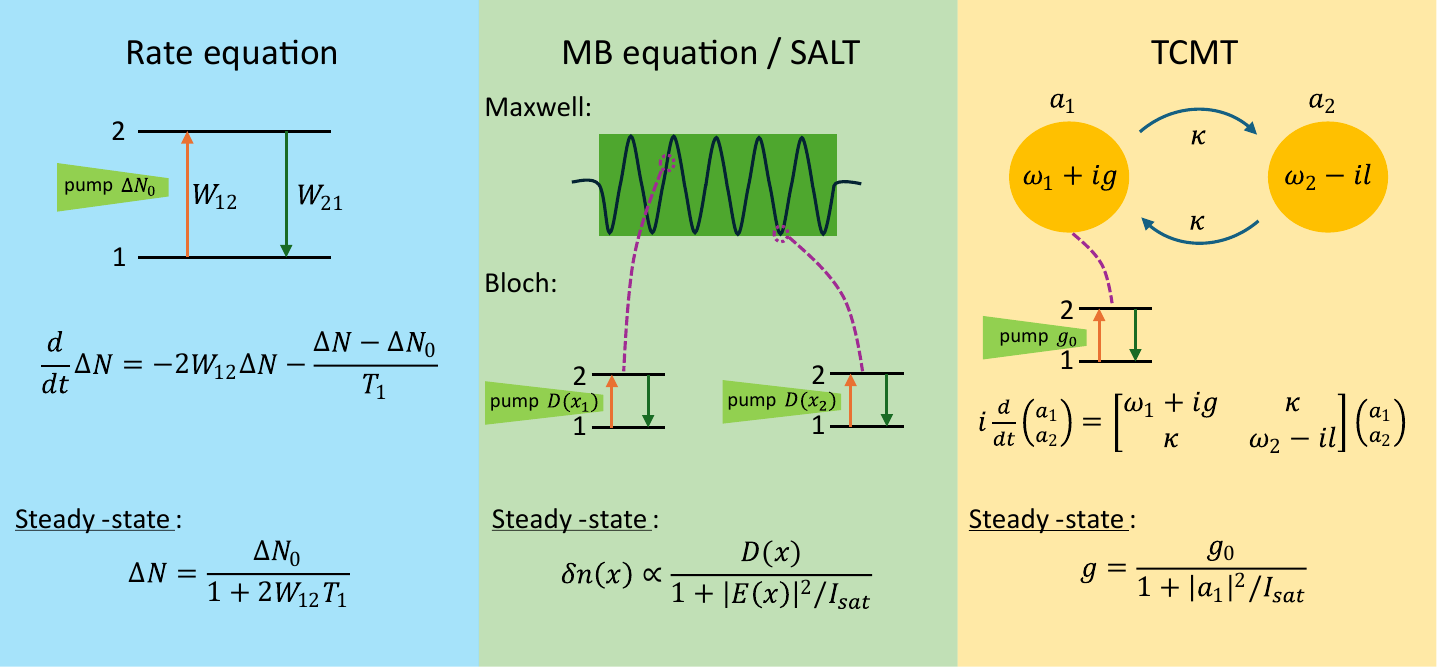}
\vspace{-6pt}
\caption{Schematic of typical laser models.
The rate equation phenomenologically describes the light-matter interaction in time while ignoring the geometry of laser cavity \cite{siegman1986lasers}.
$W_{12}$ and $W_{21}$ are the upward and downward stimulated-transition probabilities and $W_{12}\equiv W_{21}$.
$\Delta{N}=N_2-N_1$ is the population inversion and $\Delta{N}_0$ is the pumped thermal-equilibrium population inversion without external electric field.
The Maxwell-Bloch equations couple electric field with local two-level atoms, and reduce to a nonuniform perturbation on the refractive index $\delta n(x)$ caused by spatial hole burning in the static-inversion regime \cite{ge2010steady}.
Typical TCMT uses a simple saturable gain $g$ to model the complex laser dynamics \cite{hokmabadi2019non}.
}
\vspace{-4pt}
\label{fig:1}
\end{figure*}

However, if a laser is driven sufficiently close to an EP, the static gain approximation is expected to fail  \cite{gao2024dynamic,ji2023tracking}.
In this near-EP regime, the time-scale of any relaxation oscillations due to spontaneous emission %into another EP resonance 
can be sufficiently long lived that these spontaneous emission events, subsequently amplified, completely destabilize the system's single mode operation.
In such a case, even though no other cavity resonance reaches its lasing threshold, the system will still spontaneously evolve into a frequency comb with dynamic carrier populations.
%state with two permanently active frequency components.
Heuristically, the beat notes generated by oscillating inversion interacting with the lasing mode destabilizes the stationary laser gain and induces coherent oscillations of the non-resonant population inversion inside the gain medium, a process that is enhanced by the spatial coalescence of two modal profiles guaranteed by their proximity to the EP, and generating a cascade of new frequencies resulting in a self-generating frequency comb. 
However, the appearance of a self-generated frequency comb presents a problem for modeling such systems because the combined spatial-temporal complexity involves nonlinear interactions across two different time scales.
Typical laser models (Fig.~\ref{fig:1}) such as the steady-state \textit{ab initio} laser theory (SALT) and the rate equation description fail to model the dynamic gain near EP \cite{tureci2006self,ge2010steady,esterhazy2014scalable,ge2008quantitative}, because they either rely on the assumption of static population inversion (SALT) or oversimplify the geometry that is crucial for achieving the EP-condition (rate equation approach).
The only previously proven approach to correctly account for such dynamic gain is via brute-force finite-difference time-domain (FDTD) simulations of Maxwell-Bloch (MB) equations across the entire laser cavity \cite{gao2024dynamic}, which requires unusually high numerical precision due to the high sensitivity of EP to numerical error. 
Although the periodic-inversion \textit{ab initio} laser theory (PALT) analysis on MB equations provides a quantitative method for numerical computation \cite{gao2024dynamic}, it does not yield an intuitive model to explain how frequency combs are generated via local laser nonlinearities.

Here, we develop a rigorous coupled-mode analysis of the laser gain by bridging the local distribution of population inversion with the typical resonant modal interaction near an EP.
In doing so, we also demonstrate that the spatial non-uniformity induced by hole-burning in both the single-mode laser and the EP-laser can be accurately treated using a Pad{\' e} approximant.
The response of the passive cavity to the pump-induced polarization is decomposed into discrete excitations of the passive resonant modes, which is known as a quasinormal mode (QNM) expansion \cite{lalanne2018light,sauvan2022normalization,zhang2020quasinormal,binkowski2020quasinormal,alpeggiani2017quasinormal}. 
The quasinormal coupled-mode description for such EP-lasers, that we will refer to as QNM-PALT, can be expressed as integral equations with only the expansion coefficients and lasing frequencies as unknowns.
These integral equations establish the connection from local distribution of population inversion to the cavity modal interaction and quantitatively describe the enhancement of the spatial overlap between the two EP-modes with the dynamic gain.
Under the weak spatial coupling limit, which is automatically satisfied due to the EP-laser condition \cite{gao2024dynamic}, combined with the Pad{\' e} approximant of the integral function, QNM-PALT can be expressed as a purely algebraic problem of the expansion coefficients without spatial dependence.
Compared with other frameworks for modeling lasers, QNM-PALT inherits both the rigor of MB equations and conciseness of temporal coupled-mode theory (TCMT) analyses, while overcoming the mis-modeling of dynamic gain in typical phenomenological TCMT assuming static laser gain\cite{hokmabadi2019non,assawaworrarit2017robust,bai2023nonlinearity,bai2024observation}.
This simple physical model combined with its computational efficiency enables further exploration of EP lasers, including novel laser properties due to the self-time-modulation and optimization for compact source of coherent frequency combs.
The dynamic gain modeled by this QNM-PALT framework also opens new opportunities for introducing time modulation to optical systems \cite{fang2012realizing,galiffi2022photonics,sounas2017non,yuan2018synthetic}.
Moreover, the Pad{\' e} approximant we adopt here provides an efficient series expansion technique to simplify the saturable laser nonlinearity and is expected to simplify the description of gain competition in multi-mode lasers \cite{ge2010steady}.

The remainder of this paper is organized as follows:
In Sec.~\ref{sec:QNM-CMT}, we provide the theoretical background of QNM and derive the corresponding QNM expansion equations for simple 1D resonators
In Sec.~\ref{sec:single}, we apply the QNM expansion formalism to the single-mode lasing solution and introduce the Pad\' e approximant, a series expansion to simplify the field expansion integral.
In Sec.~\ref{sec:PALT}, we apply the QNM expansion method to the general frequency comb solution in EP-lasers.
We conclude in Sec.~\ref{sec:discussion}.

\section{\label{sec:QNM-CMT}Theoretical Background---Quasinormal modes and quasinormal mode expansion}
QNMs are the eigenmodes of non-Hermitian open systems.
In a 1D passive system consisting of resonators surrounded by free space and that may be absorbing, the QNMs can be defined in terms of the scalar electric field $\tilde{E}\left(x\right)$. Specifically, the $n$th QNM is defined as the solution of the source-free wave equation,
\begin{equation}
    \frac{d^2\tilde{E}_n\left(x\right)}{dx^2}+\left(\frac{\tilde{\omega}_n}{c} \right)^2\varepsilon_r\left(x\right)\tilde{E}_n\left(x\right)=0,
\end{equation}
with outgoing boundary conditions \cite{lalanne2018light,leung1994completeness,lee2009decaying}. Here,
$c$ is the light speed in vacuum and $\varepsilon_r\left(x\right)$ is the relative permittivity of the passive resonator system, which is generally complex to account for possible material absorption.
The eigenvalue $\tilde{\omega}_n=\omega_n-i\gamma_n$ is the $n$th QNM frequency and is also complex. 
Due to the causality constraint that analyticity should be maintained in the upper half plane of complex frequency \cite{Jackson:1998nia}, all QNM frequencies should have a negative imaginary part, i.e. $\gamma_n>0$.
The open boundary condition in 1D takes a simple analytical form as all waves must be outgoing plane waves $\tilde{E}_n\left(x\right)\propto e^{\pm i\tilde{k}_nx}$ beyond some surface of last scattering, where $\tilde{k}_n=\tilde{\omega}_n/c$ is the wave number in free space.
The combination of causality and the outgoing boundary condition leads to an exponentially diverging field at infinity, $\tilde{E}_n\left(x\right)\propto e^{i(\omega_n/c)|x|}e^{(\gamma_n/c)|x|}$.
Therefore, such QNMs are not square-integrable.
To overcome this difficulty and construct an orthonomal condition necessary for mode expansion, a variety of regularization schemes have been developed to mediate this divergence \cite{ren2021quasinormal,sauvan2013theory,muljarov2011brillouin,muljarov2016exact,kristensen2015normalization,ge2014design,zschiedrich2018riesz}.
In the 1D case of our interest here, it is straightforward to derive an analytic expression of its regularized inner product between QNM $\tilde{E}_n$ and $\tilde{E}_m$ as \cite{lalanne2018light,leung1994completeness,lee2009decaying}, 
\begin{multline}\label{eq:1Dnorm}
    \braket{\tilde{E}_n}{\tilde{E}_m} = \int_{x_1}^{x_2} \varepsilon_r\left(x\right) \tilde{E}_n\left(x\right)\tilde{E}_m\left(x\right)dx \\ 
    + i\frac{\tilde{E}_n\left(x_1\right)\tilde{E}_m\left(x_1\right)+\tilde{E}_n\left(x_2\right)\tilde{E}_m\left(x_2\right)}{\left(\tilde{\omega}_n+\tilde{\omega}_m\right)/c}.
\end{multline}
The unconjugated multiplication and additional boundary term come from the non-degenerate non-Hermiticity \cite{sternheim1972non} and the openness \cite{sauvan2022normalization} of this resonator system, respectively. 
The inner product in Eq.~\eqref{eq:1Dnorm} is provably unique for arbitrary choice of boundary $x_{1,2}$ in the surrounding free space and satisfies the orthogonality relation $\braket{\tilde{E}_n}{\tilde{E}_m}=\delta_{n,m}$, where $\delta_{n,m}$ is the Kronecker delta.
A detailed derivation is provided in Appendix~\ref{app:1DQNM}.

With the rigorously defined QNM inner product Eq.~\eqref{eq:1Dnorm}, a space-dependent field excitation problem can be decomposed into a few space-independent modal excitation contributions \cite{gigli2020quasinormal}. 
A resonator subject to a harmonic source excitation $P\left(x\right)$ at frequency $\omega$ generates electric field $E\left(x,t\right)$ through
\begin{equation}\label{eq:sourceexcitation}
    \frac{\partial^2E\left(x,t\right)}{\partial x^2}-\varepsilon_r\left(x\right)\frac{1}{c^2}\frac{\partial^2E\left(x,t\right)}{\partial t^2}=\frac{1}{c^2}\frac{\partial^2P\left(x\right)e^{-i\omega t}}{\partial t^2}.
\end{equation}
Expanding $E\left(x,t\right)$ onto the QNM basis as $E\left(x,t\right) = \Sigma_n a_nE_n\left(x\right)e^{-i\omega t}$, such a space-dependent partial differential equation determines a steady-state QNM amplitude as (see derivation in Appendix~\ref{app:1DQNM})
\begin{equation}\label{eq:QNMamp}
    a_n = \frac{\omega^2\int_{0}^{L}\tilde{E}_n\left(x\right)P\left(x\right)dx}{\left(\tilde{\omega}_n^2-\omega^2\right)\braket{\tilde{E}_n}{\tilde{E}_n}}.
\end{equation}
The interval of integration $\left[0,L\right]$ includes the entire resonant cavity.
In case of near resonance excitation, i.e. $|\omega-\tilde{\omega}_n|\ll|\omega|$, slowly-varying envelope approximation (SVEA) applies and the evolution of time-dependant amplitudes follows a first order rate equation akin to the conventional phenomenological temporal coupled-mode theory \cite{PhC,suh2004temporal,fan2003temporal}.
Accordingly, the frequency dependence in Eq.~\eqref{eq:QNMamp} reduces to a simple Lorentzian function.
Here we keep its original form Eq.~\eqref{eq:QNMamp} without SVEA to avoid unnecessary numerical error.

It is worth noting that the only assumption of Eq.~\eqref{eq:QNMamp} is the vanishing $P\left(x\right)$ on and outside the boundary of the cavity. 
A nonzero $P\left(x\right)$ on or outside the boundary serves as an input/output channel with a different contribution, which is discussed in \cite{zhang2020quasinormal} but not considered here. 
Therefore, such $P\left(x\right)$ can stand for different kind of source or perturbation localized in the resonator structure, such as a linear perturbation on refractive index \cite{alpeggiani2017quasinormal}, current source \cite{sauvan2013theory}, and nonlinearity \cite{gigli2020quasinormal}.

\section{\label{sec:single}Single-mode lasing under QNM expansion using the Pad\' e approximant}

In this section, we use the QNM expansion framework to investigate the single-mode lasing solution of the \textit{ab initio} MB equations and show how the Pad{\' e} approximant can be used to accurately account for spatial hole burning.
Specifically, the pump-induced polarization density $P\left(x\right)$ is considered as the perturbation and the electric field can be reconstructed by the QNM expansion.

\begin{figure}[t]
\centering
\includegraphics[width=0.5\textwidth]{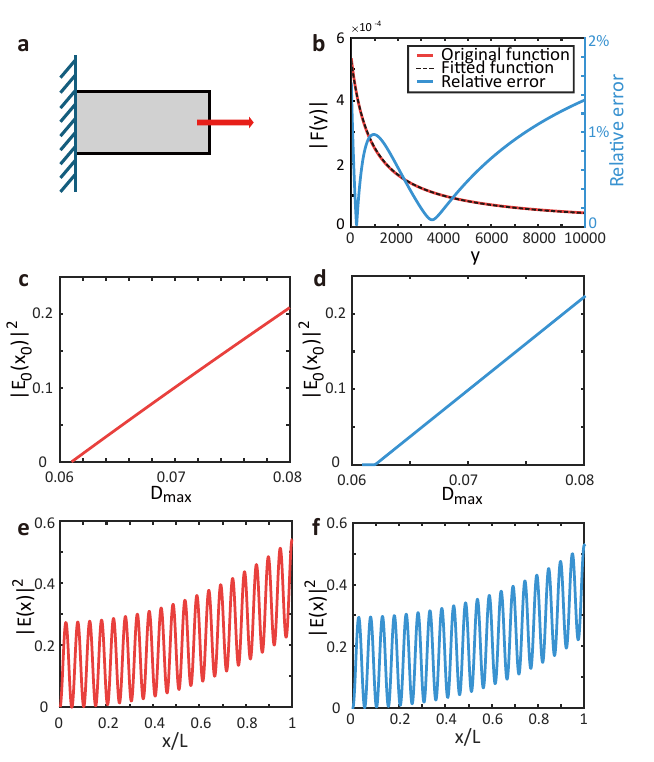}
\vspace{-6pt}
\caption{Validating the Pad{\' e} approximation using a single-mode laser in 1D cavity. 
\textbf{a} The cavity is bounded on one side by a perfect mirror and open to free space on the other side. The refractive index of the medium is $n=1.5$ and the length of the lasing medium is $L$.
Parameters of two-level atoms are $\omega_{ab}=40c/L$ and $\gamma_\perp=4c/L$.
The pump profile here is chosen uniform inside the laser cavity.
\textbf{b} Comparison of the original integral function and its Pad\' e approximant. 
Pad\' e approximant provides an accurate approximation as the relative error lies within 2\%.
\textbf{c \& d} The relation between the field intensity at certain point $x=x_0=0.24L$ away from the left mirror inside the laser cavity and the pumping strength $D_{\textrm{max}}$ computed from SALT and QNM expansion, respectively.
\textbf{e \& f} The intensity distribution across the laser cavity at $D_\textrm{max}=0.08$ from SALT and QNM expansion, respectively.
}
\vspace{-4pt}
\label{fig:2}
\end{figure}  

Consider a 1D microcavity edge emitting laser shown in Fig.~\ref{fig:2}(a) \cite{cerjan2011steady,cerjan2020modeling}.
The laser medium is modeled as an ensemble of two-level atoms and the light-matter interaction can be described semi-classically by the MB equations \cite{tureci2006self,ge2010steady}.
When subject to external pumping, a non-zero population inversion builds up and then generates a pump-induced polarization density $P\left(x\right)$ in the laser medium, providing gain to the laser cavity.
As a result, such pump-induced gain compensates the intrinsic loss of the cavity resonances gradually as the pumping strength increases, until the loss of one QNM is completely compensated.
At this point, the original complex QNM frequency $\tilde{\omega}$ shifts to a purely real lasing frequency, and this critical pumping strength is known as the first lasing threshold $D_{1}^{\textrm{th}}$.
Beyond this lasing threshold, electric field saturates the gain thus modifies the distribution of polarization nonlinearly, which, in turn, excites finite electric field from the passive cavity.
At equilibrium, a single-mode lasing operation turns on with a stationary saturated population inversion distribution, $D\left(x\right)=\left[1+|\Gamma_\perp\left(\omega_0\right)|^2|E_0\left(x\right)|^2\right]^{-1}D_\textrm{P}\left(x\right)$,
where $\Gamma_\perp\left(\omega\right)=\frac{\gamma_\perp}{\omega-\omega_{ab}+i\gamma_\perp}$ stands for the normalized gain curve and $D_\textrm{P}\left(x\right)=D_{\textrm{max}}\textrm{Win}\left(x\right)$ is the space-dependent pumping strength.
Here, $D_{\textrm{max}}$ represents the effective pumping strength and $\textrm{Win}\left(x\right)$ is a normalized window function for the pumping profile, which is zero outside the pumped cavity.
Typically, $\textrm{Win}\left(x\right)$ is not necessarily flat within the laser medium and can be tailored to be spatially nonuniform to boost the laser power-efficiency \cite{ge2014enhancement}.
$\omega_0$ and $E_0\left(x\right)$ are the real frequency and electric field determined by the nonlinear single-mode lasing equation,
\begin{widetext}
\begin{equation}\label{eq:single-SALT}
    \frac{d^2 E_0\left(x\right)}{dx^2}+\left(\frac{\omega_0}{c}\right)^2\varepsilon_r\left(x\right) E_0\left(x\right)=-\left(\frac{\omega_0}{c}\right)^2\Gamma_\perp\left(\omega_0\right)D\left(x\right)E_0\left(x\right).
\end{equation}
\end{widetext}
The intensity-dependent term in $D\left(x\right)$ accounts for the system's spatial-hole burning, which is the key to gain saturation and multi-mode lasing operation.

Under the QNM expansion framework, the left hand side of Eq.~\eqref{eq:single-SALT} representing the passive cavity response can be decomposed into modal excitation of discrete QNMs $E_0\left(x\right)=\Sigma_m a_m\tilde{E}_m\left(x\right)$.
Therefore, the spatial coupling is completely included in the QNM profile $\tilde{E}_m\left(x\right)$, leaving only the QNM-amplitudes $a_m$ and frequency $\omega_0$ as unknowns.
Ideally, the infinite set of QNMs forms a complete basis for dispersion-free permittivity $\varepsilon_r\left(x\right)$, so its result should be equivalent to directly solving Eq.~\eqref{eq:single-SALT}.
In practice, off-resonance QNM contributions decay according to a Lorentzian function $\left(\tilde{\omega}_m^2-\omega^2\right)^{-1}$, so only near-resonance QNMs dominate the cavity response.
For an excitation near a single resonance $\omega\approx\tilde{\omega}_n$, the single-mode lasing equation Eq.~\eqref{eq:single-SALT} can be simplified by the QNM expansion equation Eq.~\eqref{eq:QNMamp} with single QNM contribution $E_0\left(x\right)\approx a_n\tilde{E}_n\left(x\right)$ as \cite{ge2010steady},
\begin{equation}\label{eq:TCMTsingle}
    \left(\tilde{\omega}_n^2-\omega_0^2\right)\braket{\tilde{E}_n}{\tilde{E}_n}=\omega_0^2\Gamma_\perp\left(\omega_0\right)D_{\textrm{max}}F[|\Gamma_\perp\left(\omega_0\right)a_n|^2]
\end{equation}
where $F\left(y\right)=\int_\textrm{cav}\frac{\textrm{Win}\left(x\right)\tilde{E}_n^2\left(x\right)}{1+|\tilde{E}_n\left(x\right)|^2y} dx$ is an integral function defined across the cavity.

Equation \eqref{eq:TCMTsingle} is a nonlinear integral equation of unknowns $\{a_n,\omega_0\}$.
Typical iterative nonlinear methods for solving such an integral equation require multiple evaluations of the original function, which makes the calculation of $F\left(y\right)$ the bottleneck of computation.
Here, we find that a rational function expansion suffices to provide an accurate estimation of the integral function $F\left(y\right)$,
\begin{equation}\label{eq:pade}
    F\left(y\right)=\int_{0}^{L}\frac{\textrm{Win}\left(x\right)E_\alpha(x)^2}{1+|E_\alpha(x)|^2y} dx\approx\frac{\lambda}{1+\mu y}.
\end{equation}
This is referred to as the Pad\' e approximant of order [0/1] for function $F\left(y\right)$ \cite{baker1961pade,baker1964theory}, where the series coefficients $\{\lambda,\mu\}$ can be numerically fitted from some precomputed function values, as shown in Fig.~\ref{fig:2}(b) and discussed in Appendix ~\ref{app:Pade}.
Therefore, Eq.~\eqref{eq:TCMTsingle} for single-mode lasing reduces to a purely algebraic equation under a Pad\' e approximant Eq.~\eqref{eq:pade}.
The pump-dependent intensity and mode pattern are plotted in Fig.~\ref{fig:2}(c)-(f) for single-mode lasing solutions using the QNM expansion and direct FDFD method without approximation \cite{esterhazy2014scalable}, respectively.
By comparing results of the two methods, one can conclude that QNM expansion accurately reproduces the single-mode lasing solution.
The residual quantitative error comes from the neglect of other QNMs' excitations and the use of Pad\' e approximant. 

Notably, the nearly linear relationship between laser intensity and the pumping strength is reflected in both results in Fig.~\ref{fig:2}.
This phenomena can be readily derived from conventional interpretation of static saturable gain $g=g_0/\left(1+I/I_{\textrm{sat}}\right)$.
Stable lasing requires complete gain-loss cancellation $g=l$, thus a linear dependence $g_0=\left(1+I/I_{\textrm{sat}}\right)l$ is established between the intensity $I$ and the unsaturated gain $g_0$, which is proportional to the pumping strength $D_{\textrm{max}}$, as the loss rate $l$ can be considered constant for a given lasing mode.
However, the description of saturable gain sometimes oversimplifies the gain mechanism and cannot directly connect to the \textit{ab initio} description of MB equations, in which saturation appears as a local property known as spatial hole burning and extends the design space for laser cavities \cite{ge2014enhancement}.
Meanwhile, the linear dependence is also not straightforward from the \textit{ab initio} single-mode lasing equation Eq.~\eqref{eq:single-SALT}.

By approximating the lasing mode as a single QNM mode, one can unify these two different descriptions and provide a complete expression of the nearly linear relation.
Without external pumping, Eq.~\eqref{eq:TCMTsingle} reduces to $\omega_0=\tilde{\omega}_n$, meaning that the electric field decays at the same loss rate $l=-\textrm{Im}[\tilde{\omega}_n]$ of QNM $\tilde{E}_n(x)$.
Operating in the single-mode lasing regime means that the complex frequency change given by Eq.~\eqref{eq:TCMTsingle} has to completely cancel the intrinsic loss rate $l=-\textrm{Im}[\tilde{\omega}_n]$, leading to a real lasing frequency $\omega_0$.
As a result, by inserting Eq.~\eqref{eq:pade} into Eq.~\eqref{eq:TCMTsingle}, one can obtain a closed-form expression of the relation between laser intensity, frequency, and pumping strength,
\begin{widetext}
\begin{equation}\label{eq:linear_relation}
    1+\left[\textrm{Re}\left(\mu\right)+\frac{\textrm{Re}\left(\tilde{\omega}_n\right)^2-\textrm{Im}\left(\tilde{\omega}_n\right)^2-\omega_0^2}{2\textrm{Re}\left(\tilde{\omega}_n\right)\textrm{Im}\left(\tilde{\omega}_n\right)}\textrm{Im}\left(\mu\right)\right]|\Gamma_{\perp(\omega_0)}a_n|^2=\frac{\omega_0^2\textrm{Im}[\Gamma_{\perp}\left(\omega_0\right)\lambda]}{2\textrm{Re}\left(\tilde{\omega}_n\right)\textrm{Im}\left(\tilde{\omega}_n\right)}D_{\textrm{max}}.
\end{equation}
\end{widetext}
As pumping strength $D_{\textrm{max}}$ increases, the change in laser frequency is typically negligible so that Eq.~\eqref{eq:linear_relation} predicts a linear dependence between pumping strength $D_{\textrm{max}}$ and the laser intensity, which is proportional to $|a_n|^2$.
Therefore, QNM expansion combined with Pad\' e approximant establishes the connection between the static saturable gain and local spatial hole burning.
It is important to emphasize that the Pad\' e approximant is not merely a mathematical approximation, it is also rooted in the physics of saturable gain.
%it also implies physics of saturable gain that have been taken for granted without proof. 
The typical phenomenological description of saturable gain $g=g_0/\left(1+I/I_{\textrm{sat}}\right)$ does not include the spatial dependence, implicitly assuming a uniform distribution of both the laser medium and field intensity.
Moreover, such a description of saturable gain can successfully model many laser behaviors even though, in practice, both the pumping strength and the field intensity can be non-uniform within the laser cavity.
Here, this conflict of non-uniformity is resolved by the high accuracy of Pad\' e approximant, ensuring the effectiveness of such traditional saturable gain descriptions with the inherent spatially non-uniform laser nonlinearity.

It is worth noting that, similar linear-dependence has been found via the so-called single-pole approximation SALT (SPA-SALT) in \cite{ge2010steady}, where a different basis function is used and the lasing frequency is assumed to be fixed.
However, tracing the frequency change can be critical when a system is operating near an EP, which justifies the significance of the QNM expansion analysis here.

\section{\label{sec:PALT}Dynamic gain and frequency comb generation in near-EP lasers under QNM-PALT}

\begin{figure}[t]
\centering
\includegraphics[width=0.5\textwidth]{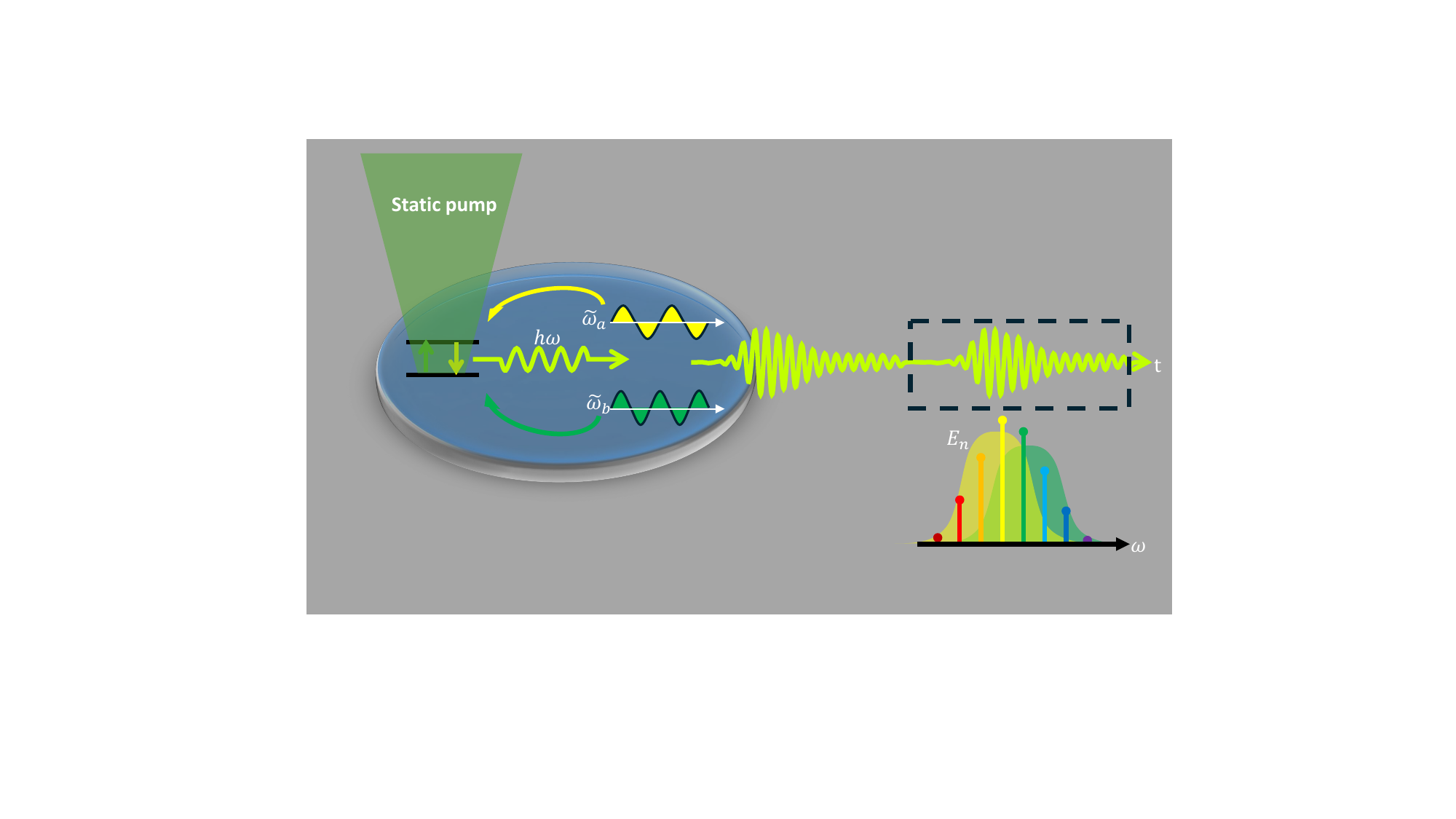}
\vspace{-6pt}
\caption{Schematic of the frequency comb generation from EP-laser.
The photons emitted from two-level atoms in the laser cavity excite both of the two passive resonant modes, the beat note of which then induces oscillating population inversion and dynamic gain.
The dynamic gain is often weak, but can be greatly enhanced by the similar field distributions of the two resonant modes realized by the EP-condition.
Modulated by such enhanced gain dynamics, the pump-induced polarization intensities excite the original two resonant modes, plotted as the two shaded area representing Lorentzian excitation line shapes, at a series of equally spaced frequencies, leading to the self-generation of frequency comb.
}
\vspace{-4pt}
\label{fig:3}
\end{figure}

Typically, adding pump power to a single-mode laser will result in a second cavity resonance reaching its lasing threshold by taking advantage of the undepleted gain at the locations in the cavity where the first mode does not have much intensity.
In this process the population inversion can remain static and yields stable two-mode lasing \cite{ge2010steady}.
However, if the system is operating close to an EP, a second frequency component can, and generally will, be populated with long-lived excitations due to spontaneous emission even when the second cavity resonance is below its lasing threshold \cite{gao2024dynamic}.
The nonlinear coupling between these two close-but-different frequencies can destabilize the stationary gain and initiate a synchronized oscillation between the population inversion and electric field.
The frequency difference determines the oscillation frequency of population inversion, which then acts as a time-modulation to the fields within the gain medium.
Thus, additional frequencies emerge through this self-time-modulation process and form a frequency comb with periodic population inversion, as illustrated in Fig.~\ref{fig:3}.
Such laser dynamics have been numerically demonstrated using PALT, a frequency domain analysis of the MB equations \cite{gao2024dynamic}.
Nevertheless, PALT also poses great challenges on both intuitive interpretation and numerical computation of dynamic gain due to the combination of nonlinear couplings across different time scales and spatial complexity.
Previously, the only feasible numerical approach for PALT relied on the fact that the response of 1D systems can be analytically described via a Green's function \cite{gao2024dynamic}, so it still requires extremely strict initial guesses and long computation time.
Here, we show that PALT can be further simplified via QNM expansion using a Pad\' e approximant to handle these difficulties.

Starting from the PALT analysis, the EP-laser dynamics  can be described by \cite{gao2024dynamic}
\begin{widetext}
\begin{align}
    &\frac{d^2E_m\left(x\right)}{dx^2}+\frac{\omega_m^2}{c^2}\varepsilon_r\left(x\right)E_m\left(x\right)=-\frac{\omega_m^2}{c^2}\Gamma_{\perp}\left(\omega_m\right)\sum_{n=-\infty}^{+\infty}D_{m-n}\left(x\right)E_n\left(x\right),\label{eq:PALTE}\\
    &\bar{D}\left(x\right)=D_\textrm{p}\left(x\right)\left[\bar{\bar{I}}- (1/2) \bar{\bar{\Gamma}}_{\parallel}\left(\bar{\bar{E}}\left(x\right)^\dag \bar{\bar{\Gamma}}_+ \bar{\bar{E}}\left(x\right)-\bar{\bar{E}}\left(x\right) \bar{\bar{\Gamma}}_-^\dag \bar{\bar{E}}\left(x\right)^\dag\right)\right]^{-1}\bar{\delta},\label{eq:PALTD}
\end{align}
\end{widetext}
where $\omega_m=\omega_0+m\omega_d$ is the frequency of the $m$th Fourier component $E_m\left(x\right)e^{-i\omega_mt}$. 
All of these Fourier components are coupled together through the dynamic population inversion $D\left(x,t\right)=\sum_n D_{n}\left(x\right)e^{-in\omega_dt}$ determined by Eq.~\eqref{eq:PALTD}, where $\bar{D}\left(x\right)$ and $\bar{\delta}$ are column vectors with components $\left(\bar{D}\left(x\right)\right)_m=D_m\left(x\right)$ and $\left(\bar{\delta}\right)_m=\delta_{m,0}$. $\delta_{m,0}$ is the Kronecker delta.
$\bar{\bar{E}}\left(x\right)$ is a matrix formed from the electric fields of each different Fourier component $\left(\bar{\bar{E}}\left(x\right)\right)_{mn}=E_{m-n}\left(x\right)$.
$\bar{\bar{\Gamma}}_{\parallel}$ and $\bar{\bar{\Gamma}}_{\pm}$ are diagonal matrices with $\left(\bar{\bar{\Gamma}}_{\parallel}\right)_{mn}=\delta_{m-n}\gamma_\parallel/\left(m\omega_d+i\gamma_\parallel\right)$ and $\left(\bar{\bar{\Gamma}}_{\pm}\right)_{mn}=\delta_{m-n}\Gamma_\perp\left(\omega_{\pm m}\right)$, representing the dispersive response of the two-level atoms. 
$\gamma_\parallel$ is the relaxation rate of population and
$\bar{\bar{I}}$ is the identity matrix.
Eq.~\eqref{eq:PALTD} describes how a static pump $D_\textrm{p}\left(x\right)$ is saturated and modulated locally by the electric field $E\left(x\right)=\sum_m E_{m}\left(x\right)e^{-i\omega_mt}$, and then the induced polarization density serves as an excitation source to support the stable generation of frequency comb through Eq.~\eqref{eq:PALTE}.

\begin{figure*}[t]
\centering
\includegraphics[width=0.8\textwidth]{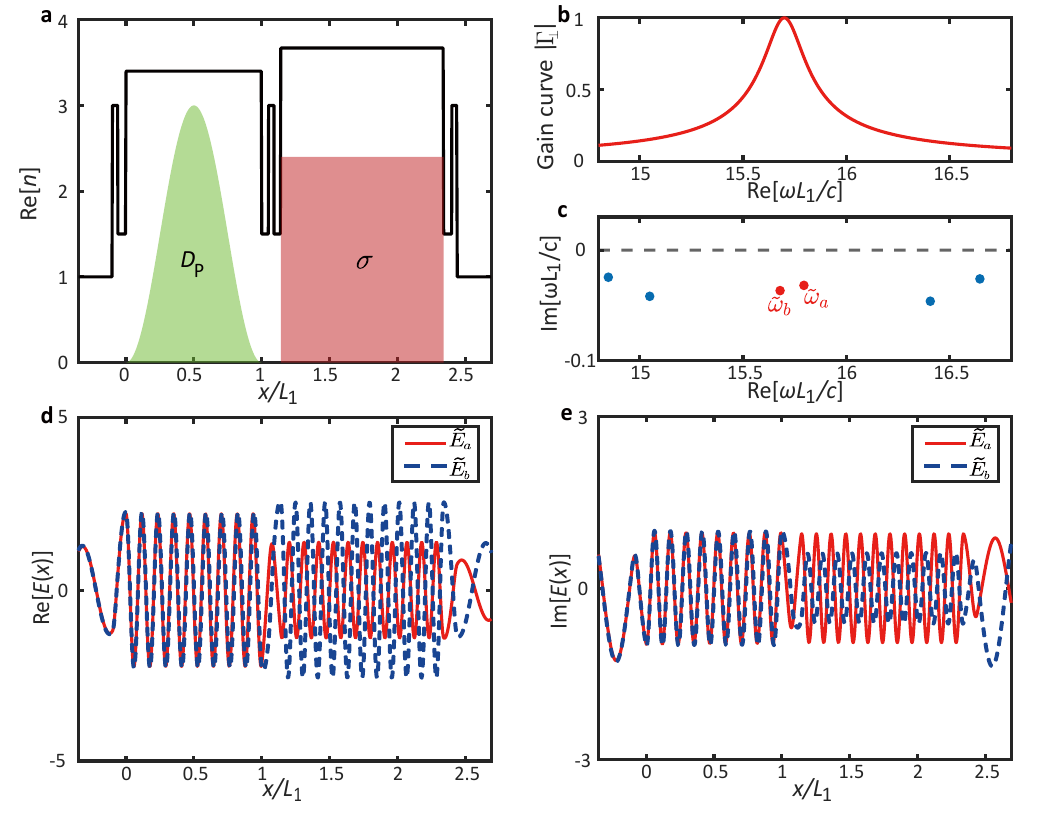}
\vspace{-6pt}
\caption{
    \textbf{a}, The refractive index profile of the coupled gain-loss resonator.
    The green area stands for the pump profile and the red area represents the passive conductivity $\sigma$. 
    The pump profile is tapered to a Hann window function $\textrm{Win}\left(x\right)=1-\textrm{cos}\left(2\pi x/L_\textrm{1}\right)$ within the active cavity and the gain parameters are $\gamma_\perp=0.1 c/L_1$ and $\gamma_\parallel=0.002c/L_1$.
    The conductivity $\sigma$ is tuned when increasing pumping strength to ensure the laser operates near EP.
    \textbf{b}, Absolute value of the normalized gain curve $\Gamma_\perp(\omega)$. 
    \textbf{c}, Distribution of QNM frequencies on the complex frequency plane. The two modes around the center of gain curve have biggest excitation amplitude thus are colored red. 
    \textbf{d}-\textbf{e}, The real and imaginary part of the QNM field profile $\tilde{E}_{a,b}$ across the resonator. These two modes share similar profile inside the two cavities with a constant phase different.}
\vspace{-4pt}
\label{fig:4}
\end{figure*}

For illustrative purposes, we consider a parity-time-symmetric-like configuration as shown in Fig.~\ref{fig:4}\textbf{a}, where the pumped cavity is coupled to the passive cavity with linear material loss.
The real part of refractive indexes of the pumped and lossy cavities are $n_1=3.4$ and $n_2=3.67$.
The system is normalized by the length of pumped cavity $L_1=2.05\mu m$, and the length of the lossy cavity is $L_2=1.2L_1$.
Two DBRs are used at both ends of the structure to control the radiation loss, and its corresponding refractive indexes and lengths are $\left[n=3.0, L=0.04L_1\right]$ and $\left[n=1.5, L=0.06L_1\right]$.
The DBR in the middle regulates the coupling between the two cavity modes, and its refractive indexes and lengths are $\left[n=3.0, L=0.04L_1\right]$ and $\left[n=1.5, L=0.05L_1\right]$.
The active cavity $\textrm{C}_{\textrm{act}}$ on the left is immersed in a tapered static pump $D_\textrm{P}\left(x\right)$ and the passive cavity $\textrm{C}_{\textrm{pas}}$ on the right has linear absorption in terms of a complex permittivity $\varepsilon_r\left(x\right)=n_2^2+i\sigma/\omega$, where $\sigma$ is the passive conductivity that yields absorption.
The pump provides dispersive gain to signals at different frequencies, and the corresponding modulus of normalized gain curve $\Gamma_\perp\left(\omega\right)=\frac{\gamma_\perp}{\omega-\omega_{ab}+i\gamma_\perp}$ is shown in Fig.~\ref{fig:4}\textbf{b}.
Such a coupled-cavity structure has its QNM frequencies located below the real axis as shown in Fig.~\ref{fig:4}\textbf{c}.
The two QNMs around the center of the gain curve receive most gain compared with other QNMs, thus become the EP pair we will focus on through this section. The field profiles of the EP pair are depicted across the coupled-cavity system in
Figs.~\ref{fig:4}\textbf{d}-\textbf{e}, which show similar field profiles within each cavity and a cross-cavity tunneling due to the weak spatial coupling suppressed by the DBR in the middle.

\begin{figure}[t]
\centering
\includegraphics[width=0.5\textwidth]{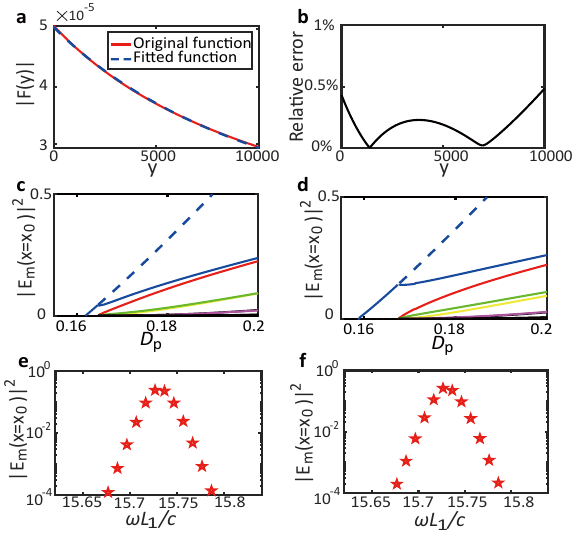}
\vspace{-6pt}
\caption{Comparison of computation results from exact Green's function brute-force integration and QNM-PALT.
\textbf{a} Absolute value of the original integral function $F(y)$ and the fitted function with real variable y.
\textbf{b} Percentage error of the Pad\' e approximant compared with the exact value.
\textbf{c\&d} Lasing intensities at frequency $\omega_m=\omega_0+m\omega_d$ as a function of pumping strength $D_{\textrm{max}}$.
The level of absorption $\sigma$ is tuned with the pump to ensure near-EP operation.
\textbf{e\&f}, The comb spectrum at pumping strength of $D_{\textrm{max}}=0.2$.
\textbf{c\&e} are the exact result from Green's function and \textbf{d\&f} are from QNM-PALT under Pad\' e approximant.
}
\vspace{-4pt}
\label{fig:5}
\end{figure}

Solving the spatial-temporally coupled nonlinear wave equations Eq.~\eqref{eq:PALTE}-\eqref{eq:PALTD} reveals the frequency comb formation in this near-EP system.
Its exact solution is plotted in Fig.~\ref{fig:5}\textbf{c} as a function of $D_\textrm{max}$, with each line representing the intensity $|E_m\left(x_0\right)|^2$ of a comb tooth at the left end of the active cavity $\textrm{C}_{\textrm{act}}$ $x_0=0$.
The EP comb spectrum at pumping strength of $D_\textrm{max}=0.2$ is plotted in Fig.~\ref{fig:5}\textbf{e}.
Only the central two comb teeth with highest intensity can be continuously connected to the two original passive QNMs, while the rest come from wave mixing induced by the dynamic population inversion.
However, since the temporal coupling terms in Eq.~\eqref{eq:PALTE}-\eqref{eq:PALTD} include nonlinear matrix inverse which should be evaluated at every spatial pixel, under standard PALT one has to solve a nonlinear dense matrix problem with a massive number of discretized field as unknowns.
Although the simplicity of a 1D structure allows a semi-analytical Green's function approach for numerical computation \cite{gao2024dynamic}, it is generally impractical to implement in more complicated structures in 2D and 3D.

Instead, under the QNM expansion framework, the spatial dependence of PALT equations can be included in the QNM basis, leaving only a few temporally coupled amplitudes as unknowns. 
Moreover, as the pump-induced polarization in Eq.~\eqref{eq:PALTE} is bounded by the bandwidth of the gain curve shown in Fig.~\ref{fig:4}\textbf{b}, the two near-resonance QNM excitations marked as red dots in Fig.~\ref{fig:4}\textbf{c} dominate the passive cavity response, even though the other comb lines generated by frequency-mixing of the original two frequencies are not directly connected to these two QNMs.
Additionally, we notice that these two QNMs shown in Fig.~\ref{fig:4} \textbf{d-e} have almost the same field spatial profile $E_\alpha\left(x\right)$ inside the pumped cavity $\tilde{E}_{a,b}\left(x\right)=\alpha_{1,2}E_\alpha\left(x\right)$, guaranteed by the weak spatial coupling limit in such EP-laser \cite{gao2024dynamic}.
By projecting the electric field onto the near-resonance 2-QNM basis as $E_m\left(x\right)=a_m\tilde{E}_a\left(x\right)+b_m\tilde{E}_b\left(x\right)$ and adopting the same profile approximation, Eq.~\eqref{eq:PALTE} can be simplified as a set of nonlinear matrix integral equations with the QNM amplitudes $\left\{a_m,b_m\right\}$ and frequencies $\left\{\omega_0,\omega_d\right\}$ as unknowns (see detailed derivation in Appendix ~\ref{app:PALT}),
\begin{widetext}
\begin{align}
    &\left(\tilde{\omega}_a^2-\omega_m^2\right)a_m\braket{\tilde{E}_a}{\tilde{E}_a}=\omega_m^2D_{\textrm{max}}\alpha_1\sum_kM_{m-k}\left(a_k\alpha_1+b_k\alpha_2\right),\label{eq:TCMTcomb_A}\\
    &\left(\tilde{\omega}_b^2-\omega_m^2\right)b_m\braket{\tilde{E}_b}{\tilde{E}_b}=\omega_m^2D_{\textrm{max}}\alpha_2\sum_kM_{m-k}\left(a_k\alpha_1+b_k\alpha_2\right),\label{eq:TCMTcomb_B}
\end{align}
\end{widetext}
where $M_{m-k}$ is the coupling coefficient between signals with frequency $\omega_m$ and $\omega_k$ induced by the $\left(m-k\right)$th component $D_{m-k}$ of dynamic population inversion.
Inserting the QNM expansion into Eq.~\eqref{eq:PALTD}, one can determine $M_{m-k}$ by $M_{m-k}=\left(\bar{M}\right)_{m-k}$, where $\bar{M}$ is a column vector,
\begin{equation}\label{eq:Mvector}
    \bar{M}=\bar{\bar{F}}\left(\bar{\bar{I}}_{\textrm{eff}}\right)\bar{\delta}.
\end{equation}
Here, $\bar{\bar{F}}\left(\bar{\bar{I}}_{\textrm{eff}}\right)$ is a matrix generalization of the scalar function $F\left(y\right)$ defined as $[\bar{\bar{F}}\left(\bar{\bar{I}}_{\textrm{eff}}\right)]_{mn}=F[(\bar{\bar{I}}_{\textrm{eff}})_{mn}]$. 
$\bar{\delta}$ is a column vector with components $\left(\bar{\delta}\right)_m=\delta_{m,0}$.
$\bar{\bar{I}}_{\textrm{eff}}$ is an effective intensity matrix defined as
\begin{eqnarray}
    \bar{\bar{I}}_{\textrm{eff}}=\frac{1}{2}\bar{\bar{\Gamma}}_\parallel&\left[\left(\alpha_1\bar{\bar{a}}+\alpha_2\bar{\bar{b}}\right)\bar{\bar{\Gamma}}_-^\dag\left(\alpha_1^*\bar{\bar{a}}^\dag+\alpha_2^*\bar{\bar{b}}^\dag\right)\right.\nonumber \\
     -&\left.\left(\alpha_1^*\bar{\bar{a}}^\dag+\alpha_2^*\bar{\bar{b}}^\dag\right)\bar{\bar{\Gamma}}_+\left(\alpha_1\bar{\bar{a}}+\alpha_2\bar{\bar{b}}\right)\right],\label{eq:Ieff}
\end{eqnarray}
where $\left(\bar{\bar{a}}\right)_{mn}=a_{m-n}$ and $\left(\bar{\bar{b}}\right)_{mn}=b_{m-n}$ are two amplitude matrices.

Equations~\eqref{eq:TCMTcomb_A}-\eqref{eq:Ieff} are the basic QNM-PALT integral equations of the QNM amplitudes $\left\{a_m,b_m\right\}$ and frequencies $\left\{\omega_0,\omega_d\right\}$.
However, elements of $\bar{\bar{I}}_{\textrm{eff}}$ are generally complex and might fall outside the valid domain of Pad\' e apprixmant, which means Pad\' e apprixmant cannot be adopted directly to simplify the integral calculation.
Noting that $\bar{\bar{I}}_{\textrm{eff}}$ only depends on the QNM amplitudes and has no spatial dependence, its eigenvalues and eigenvectors should also be independent of space.
Therefore, its matrix of eigenvectors can be taken out of the integral thus significantly reducing the number of evaluations of $F\left(y\right)$ via 
\begin{equation}
    \bar{M}=\bar{\bar{F}}\left(\bar{\bar{I}}_{\textrm{eff}}\right)\bar{\delta}
    =\bar{\bar{F}}\left(\bar{\bar{P}}^{-1}\bar{\bar{\Lambda}}\bar{\bar{P}}\right)\bar{\delta}
    =\bar{\bar{P}}^{-1}\bar{\bar{F}}\left(\bar{\bar{\Lambda}}\right)\bar{\bar{P}}\bar{\delta}
\end{equation}
where $\bar{\bar{P}}$ and $\bar{\bar{\Lambda}}$ are defined by the diagonalization of $\bar{\bar{I}}_{\textrm{eff}}=\bar{\bar{P}}^{-1}\bar{\bar{\Lambda}}\bar{\bar{P}}$.
One only needs to evaluate $F\left(y\right)$ at the diagonal elements since $\bar{\bar{\Lambda}}$ is the diagonal eigenvalue matrix.
Moreover, as evidenced in Appendix ~\ref{app:Pade}, all the eigenvalues in $\bar{\bar{\Lambda}}$ reside within the valid domain of Pad\' e apprixmant, so all the integrals can be further simplified by Pad\' e apprixmant accurately.
As a result, the QNM-PALT equations Eq.~\eqref{eq:TCMTcomb_A}-\eqref{eq:Ieff} reduce to purely algebraic nonlinear equations without any spatial dependence or integration.

As illustrated in Fig.~\ref{fig:5}\textbf{a}-\textbf{b}, this Pad\' e approximant provides a rather accurate estimation of $F(y)$ for real $y$ and can be analytically continued to a finite area on the complex $y$ plane (a more detailed discussion about the analytic continuation and its accuracy is provided in Appendix ~\ref{app:Pade}).
Solving these nonlinearly coupled algebraic QNM-PALT equations yields numerical results shown in Fig.~\ref{fig:5}\textbf{d} and \textbf{f}.
The corresponding comb spectrum is quantitatively reproduced by our QNM-PALT formalism.
The residual quantitative error comes from the neglect of other QNMs' excitations and the use of Pad\' e approximant.

The original PALT equations Eq.~\eqref{eq:PALTE}-\eqref{eq:PALTD} define a matrix problem sparsely coupled in space but densely coupled in frequency.
Direct solution approaches such as finite-difference frequency domain (FDFD) are impractical due to the small discretization size required and high density of this matrix problem \cite{esterhazy2014scalable}.
Although a Green's function can analytically describe spatial dependence in Maxwell equation for 1D systems, the spatially non-uniform population inversion still requires fine spatial discretization and restricts the Green's function approach to simple 1D structures \cite{gao2024dynamic}.
Instead, the QNM-PALT we develop here completely eliminates the space dependence in PALT and provides an accurate description of the coupling strength among different frequencies.
The issue of discretizing non-uniform population inversion is circumvented by reconstruction from the QNM expansion coefficients and the field profile of QNM, so spatial discretization on population inversion is no longer needed in QNM-PALT.
Furthermore, the QNM-PALT formalism does not depend on the physical dimension of the system and can therefore be directly generalized to 2D and 3D structures---one just need to replace the 1D QNM inner product Eq.~\eqref{eq:1Dnorm} with the form of the practical dimension \cite{lalanne2018light}.

\section{\label{sec:discussion}Discussion}

The dynamic distribution of population inversion in near-EP lasers has been fully described by the PALT theory \cite{gao2024dynamic}, which predicts the self-generation of a frequency comb near an EP.
However, it is still challenging to numerically model the coupling between local laser nonlinearity and resonant modal interaction of EP, especially in 2D and 3D.
The QNM-PALT framework we developed here provides a systematic approach to bridge these two processes at different time scales.
Unlike traditional phenomenological TCMT, which is constructed empirically and might require numerical experiments to fit key parameters such as coupling coefficients and decaying rates \cite{rodriguez2007chi,burgess2009difference,wang2020petermann,ji2023tracking}, QNM-PALT utilizes the rigorously defined QNMs as the basis functions to reconstruct the response of the passive structure to an external laser pump.
All key parameters in QNM-PALT have closed form expressions and can be computed with passive QNM solutions \cite{lalanne2018light}.
Altogether, by inheriting both the conciseness of TCMT and the rigor of \textit{ab initio} MB equations, QNM-PALT provides a minimal model to describe the dynamic gain in EP-lasers, in which many interesting EP behaviors realized on other nonlinear platforms can hopefully be revisited in the context of laser nonlinearity \cite{wang2019dynamics,bai2023nonlinearity,bai2023nonlinear,bai2024observation,ramezanpour2021tuning,pick2017enhanced,assawaworrarit2017robust}.
Moreover, the spatial degree of freedom in laser design can also be efficiently investigated using our approach \cite{ge2014enhancement}.

Under reasonable approximations, the challenging numerical computation of PALT can be further simplified to the set of algebraic equations comprising QNM-PALT and can be solved with minimal effort using ordinary nonlinear system solvers.
In particular, the Pad\' e approximant of the expansion integral function provides an accurate numerical approach to eliminate the spatial dependence in both EP-lasers and typical single-mode lasers.
Compared with a regular Taylor series expansion \cite{gigli2020quasinormal}, such a rational function expansion is more consistent with the characteristics of gain saturation and is expected to simplify the spatial dependence in more complicated multi-mode lasers.
The computational simplicity QNM-PALT promises makes it possible for generalizing such EP-lasers to 2D and 3D structures commonly used in on-chip devices, where more degrees of freedom are available for fine tuning towards EP \cite{ruter2010observation,peng2014parity,peng2014loss,brandstetter2014reversing,feng2014single,hodaei2014parity}.
Moreover, our approach also enables optimization of the laser cavity to produce high-quality coherent frequency combs with a compact device volume.

\begin{acknowledgments}
A.C.\ acknowledges support from the Laboratory Directed Research and Development program at Sandia National Laboratories.
This work was performed in part at the Center for Integrated Nanotechnologies, an Office of Science User Facility operated for the U.S. Department of Energy (DOE) Office of Science.
Sandia National Laboratories is a multimission laboratory managed and operated by National Technology \& Engineering Solutions of Sandia, LLC, a wholly owned subsidiary of Honeywell International, Inc., for the U.S. DOE's National Nuclear Security Administration under Contract No. DE-NA-0003525. 
The views expressed in the article do not necessarily represent the views of the U.S. DOE or the United States Government.
\end{acknowledgments}

\appendix

\section{\label{app:1DQNM}Quasinormal mode and quasinormal expansion in 1D}
In this section, we first introduce the concept of QNM and its corresponding properties in 1D systems, then derive the QNM expansion equation Eq.~\eqref{eq:QNMamp}.

For a 1D resonator in free space environment, QNMs are defined as the solution to source-free eigenvalue problem
\begin{equation}\label{eq:waveequation}
    \frac{d^2\tilde{E}_m\left(x\right)}{dx^2}+\left(\frac{\tilde{\omega}_m}{c}\right)^2\varepsilon_r\left(x\right)\tilde{E}_m\left(x\right)=0,
\end{equation}
with outgoing boundary condition $\tilde{E}_m\left(x=x_{1,2}\right)\propto e^{\mp i\left(\tilde{\omega}_m/c\right)x}$, where $\{x_1,x_2\}$ are two arbitrary points in the uniform environment.

Due to the radiation loss and material absorption, all QNM frequencies in this resonator should have negative imaginary part.
The interplay of outgoing boundary condition and the negative imaginary part leads to an exponentially diverging tail of the QNM field outside the resonator.
Therefore, QNM profiles extend to infinity and are intrinsically not square-integrable.

Instead of integrating the wave function in the infinite space, it is possible to integrate the wave function within a finite region enclosed by the two points $x=x_{1,2}$ in the surrounding free space.
Multiply Eq.~\eqref{eq:waveequation} by $\tilde{E}_{n\left(x\right)}$ then integrate over region $[x_1,x_2]$, we get
\begin{equation}\label{eq:finiteintegrate}
    \int_{x_1}^{x_2}\left[\tilde{E}_n\left(x\right)\frac{d^2\tilde{E}_m\left(x\right)}{dx^2}+\frac{\tilde{\omega}_m^2}{c^2}\varepsilon_{r}\left(x\right)\tilde{E}_n\left(x\right)\tilde{E}_m\left(x\right)\right]dx=0.
\end{equation}
Integrating by parts, Eq.~\eqref{eq:finiteintegrate} can be written as
\begin{widetext}
\begin{equation}
    \left[\tilde{E}_n\left(x\right)\frac{d\tilde{E}_m\left(x\right)}{dx}-\tilde{E}_m\left(x\right)\frac{d\tilde{E}_n\left(x\right)}{dx}\right]_{x_1}^{x_2}+\int_{x_1}^{x_2}\tilde{E}_m\left(x\right)\frac{d^2\tilde{E}_n\left(x\right)}{dx^2}dx+\frac{\tilde{\omega}_m^2}{c^2}\int_{x_1}^{x_2}\varepsilon_{r}\tilde{E}_n\left(x\right)\tilde{E}_m\left(x\right)dx=0.
\end{equation}
\end{widetext}
Subtracting $\int_{x_1}^{x_2}dx~\tilde{E}_m\left(x\right)\cdot$ Eq.~\eqref{eq:waveequation} with $m\rightarrow n$, then substituting the first derivative using the outgoing boundary condition $\frac{d\tilde{E}_{m}\left(x_{1,2}\right)}{dx}=\mp i\frac{\tilde{\omega}_m}{c}\tilde{E}_{m}\left(x_{1,2}\right)$, we obtain the orthogonality relation,
\begin{widetext}
\begin{equation}
    \label{eq:1Dorthogonality}
    \left(\tilde{\omega}_m-\tilde{\omega}_n\right)\left[i\frac{\tilde{E}_{m}\left(x_{1}\right)\tilde{E}_{n}\left(x_{1}\right)+\tilde{E}_{m}\left(x_{2}\right)\tilde{E}_{n}\left(x_{2}\right)}{\left(\tilde{\omega}_m+\tilde{\omega}_n\right)/c}+\int_{x_1}^{x_2}\varepsilon_r \tilde{E}_n\left(x\right)\tilde{E}_m\left(x\right)dx\right]=0.
\end{equation}
\end{widetext}
The term in the bracket is defined as the regularized inner product $\langle \tilde{E}_n|\tilde{E}_m \rangle$. 
Obviously, in non-degenerate systems, $\langle \tilde{E}_n|\tilde{E}_m \rangle=0$ for $n\neq m$ and $\langle \tilde{E}_n|\tilde{E}_m \rangle$ can be nonzero only when $n=m$, which defines the normalization of $\tilde{E}_m$.

It is worth noting that, although orthogonality relation Eq.~\eqref{eq:1Dorthogonality} is derived by integrating from $x_1$ to $x_2$, such defined inner product is unique and independent of the choice of $\left\{x_1,x_2\right\}$, so long as $\left\{x_1,x_2\right\}$ are both chosen in the uniform environment medium.
The proof is straightforward as the change in the surface term caused by different choice of $\left\{x_1',x_2'\right\}$ cancels the change in the volume integral.
Therefore, this inner product is unique and carries physical significance for the QNMs.

With the inner product defined above, consider a harmonic polarization density $P(x,t)=P\left(x\right)e^{-i\omega t}$ as the source to excite the resonator. 
The excited field $E(x,t)$ is determined by
\begin{equation}\label{eq:Pexcitation}
    \frac{\partial^2E\left(x,t\right)}{\partial x^2}-\varepsilon_r\left(x\right)\frac{1}{c^2}\frac{\partial^2E\left(x,t\right)}{\partial t^2}=\frac{1}{c^2}\frac{\partial^2\left[P\left(x\right)e^{-i\omega t}\right]}{\partial t^2}.
\end{equation}
Expanding the electric field onto the QNM basis as $E\left(x,t\right)=\sum_n A_n\left(t\right)E_n\left(x\right)e^{-i\tilde{\omega}_nt}$, where $A_n\left(t\right)$ is the time-dependent amplitude of QNM $\tilde{E}_n$.
Substitute it back into Eq.~\eqref{eq:Pexcitation}, we obtain
\begin{widetext}
\begin{equation}
    \sum_n-\varepsilon_r\tilde{E}_n\left(x\right)\frac{1}{c^2}\left(\frac{d^2A_n}{dt^2}-2i\tilde{\omega}_n\frac{dA_n}{dt}\right)e^{-i\tilde{\omega}_nt}=-\frac{\omega^2}{c^2}P\left(x\right)e^{-i\omega t}.
\end{equation}
\end{widetext}
For near resonance excitation $|\omega-\tilde{\omega}_m|\ll|\omega|$, the second derivative terms on the left side is small relative to the first derivatives and thus can be ignored.
This is known as the slowly-varying envelope approximation (SVEA).
By taking the inner product of both sides with ${\tilde{E}_m}$, we obtain
\begin{equation}
    \frac{dA_m\left(t\right)}{dt}=-\frac{\omega^2\left[\int_{x_1}^{x_2}\tilde{E}_m\left(x\right)P\left(x\right)dx\right]}{2i\tilde{\omega}_m\langle \tilde{E}_m|\tilde{E}_m \rangle}e^{-i\left(\omega-\tilde{\omega}_m\right)t}.
\end{equation}
Here, $P\left(x\right)$ is assumed to be zero outside the resonator structure.
After a gauge transform: $a_m\left(t\right)=A_m\left(t\right)e^{-i\tilde{\omega}_m t}$, the QNM expansion equations take the form of typical TCMT,
\begin{equation}\label{eq:QNM-TCMT}
    \frac{da_m\left(t\right)}{dt}=-i\tilde{\omega}_ma_m\left(t\right)-\frac{\omega^2\left[\int_{x_1}^{x_2}\tilde{E}_m\left(x\right)P\left(x\right)dx\right]}{2i\tilde{\omega}_m\langle \tilde{E}_m|\tilde{E}_m \rangle}e^{-i\omega t}.
\end{equation}
At steady state, the field would synchronize with the harmonic source and oscillate at the same frequency: $a_m\left(t\right)=a_me^{-i\omega t}$.
The steady state amplitude of QNM $\tilde{E}_m$ is thus given by
\begin{equation}\label{eq:QNMamplitudeSVEA}
    a_m=\frac{\omega^2\left[\int_{x_1}^{x_2}\tilde{E}_m\left(x\right)P\left(x\right)dx\right]}{2\tilde{\omega}_m\left(\tilde{\omega}_m-\omega\right)\langle \tilde{E}_m|\tilde{E}_m \rangle}.
\end{equation}
Or if SVEA not adopted
\begin{equation}\label{eq:QNMamplitude}
    a_m=\frac{\omega^2\left[\int_{x_1}^{x_2}\tilde{E}_m\left(x\right)P\left(x\right)dx\right]}{\left(\tilde{\omega}_m^2-\omega^2\right)\langle \tilde{E}_m|\tilde{E}_m \rangle}.
\end{equation}

Similar to the 1D case discussed above, QNMs in 3D are also defined as the eigensolution of the source-free Maxwell equations with outgoing boundary conditions.
The corresponding inner product in 3D can be obtained from the unconjugated form of the Lorentz reciprocity theorem \cite{yan2018rigorous}.
In practice, outgoing boundary conditions are usually implemented with perfectly matched layers (PMLs), which can damp out the exponential growth of QNMs away from the resonator and make QNMs square-integrable \cite{sauvan2013theory}.
The vanishing field at the outer surface of PMLs leads to a zero surface integral in the unconjugated Lorentz reciprocity theorem, and thus defines a complete and orthogonal basis formed by QNMs.
The orthogonal inner product defined with PMLs suffices to determine the expansion coefficients \cite{yan2018rigorous,gigli2020quasinormal}.

\section{\label{app:belowD1}QNM expansion: below first lasing threshold}
In the following sections, the QNM expansion formalism is applied to the PALT theory.
Here we consider the same 1D coupled-cavity resonator structure described in the main text.

\begin{figure}
    \centering
    \includegraphics[width=1.05\linewidth]{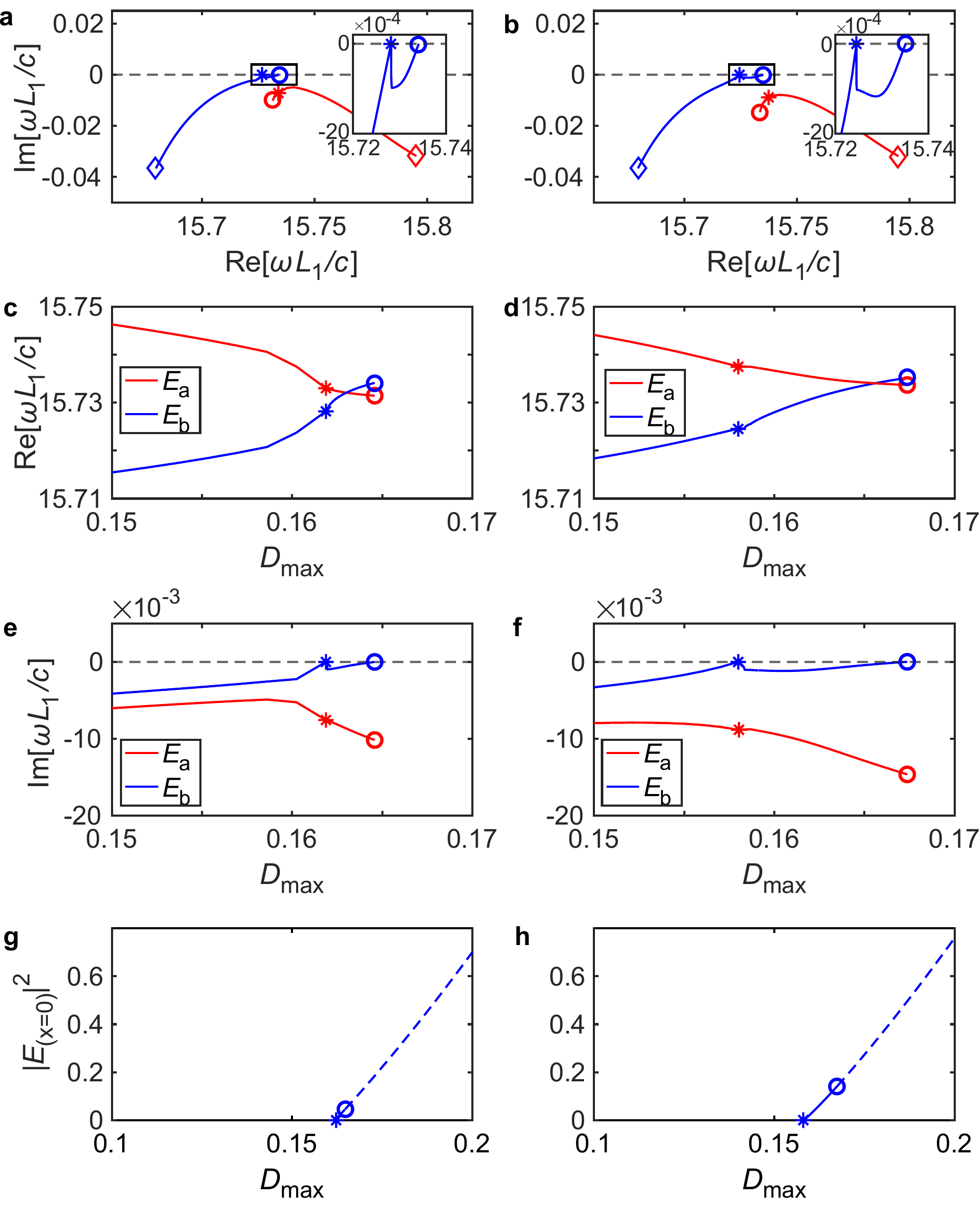}
    \caption{
    The left column is the exact result from Green's function, and the right column is the result from QNM expansion.
    \textbf{a}-\textbf{b}, The trajectory of the complex perturbation frequency as pumping strength $D_{\textrm{max}}$ increases.
    The inset shows a zoom-in plot of the boxed area.
    \textbf{c}-\textbf{f}, The evolution of complex perturbation frequency as a function of pumping strength $D_{\textrm{max}}$.
    The square, star, and circle represent $D_{\textrm{max}}=0$, $D_{\textrm{max}}=D_1^{th}$, $D_{\textrm{max}}=D_c^{th}$, respectively.
    $D_1^{th}=0.1619, D_c^{th}=0.1646$ for Green's function result and $D_1^{th}=0.1580, D_c^{th}=0.1674$ for QNM expansion result.
    \textbf{g}-\textbf{h}, Single-mode lasing intensity as a function of pumping strength $D_{\textrm{max}}$.
    Dashed lines correspond to unstable solutions.
    }
    \label{fig:stability}
\end{figure}

When the active medium is pumped at a relatively small pumping strength $D_{\textrm{max}}$, the pump-induced polarization density $P\left(x\right)=\Gamma_\perp\left(\omega\right) D_\textrm{P}\left(x\right)E\left(x\right)$ can be treated as an external driving source. 
Here $\Gamma_\perp\left(\omega\right)=\gamma_\perp/\left(\omega-\omega_{ab}+i\gamma_\perp\right)$ is the dispersive gain curve.
$D_\textrm{P}\left(x\right)=D_{\textrm{max}}\textrm{Win}\left(x\right)$ is the pump distribution, $D_{\textrm{max}}$ is the pumping strength and $\textrm{Win}\left(x\right)$ is the fixed pump profile.
Since the excited amplitudes Eq.~\eqref{eq:QNMamplitude} have a dispersive contribution which is maximized when the frequency difference is small, considering the contribution from the two modes near the gain center is sufficient to model this lasing behavior.
So, the electric field can be expanded onto this 2-QNM basis as $E\left(x\right)\approx a\tilde{E}_a\left(x\right)+b\tilde{E}_b\left(x\right)$ and $\left\{a,b\right\}$ are the constant QNM amplitudes.

Under the QNM expansion, the wave equation with $P\left(x\right)=\Gamma_\perp\left(\omega\right) D_\textrm{P}\left(x\right)E\left(x\right)$ as a source can be written as
\begin{widetext}
\begin{align}
    &\left(\tilde{\omega}_a^2-\omega^2\right)a\langle \tilde{E}_a|\tilde{E}_a \rangle=\omega^2\Gamma_\perp\left(\omega\right)D_{\textrm{max}}\left[\left(\int_{x_1}^{x_2}\textrm{Win}\left(x\right)\tilde{E}_a\left(x\right)^2dx\right)a+\left(\int_{x_1}^{x_2}\textrm{Win}\left(x\right)\tilde{E}_a\left(x\right)\tilde{E}_b\left(x\right)dx\right)b\right],\label{TCMTlinear_A}\\
    &\left(\tilde{\omega}_b^2-\omega^2\right)b\langle \tilde{E}_b|\tilde{E}_b \rangle=\omega^2\Gamma_\perp\left(\omega\right)D_{\textrm{max}}\left[\left(\int_{x_1}^{x_2}\textrm{Win}\left(x\right)\tilde{E}_b\left(x\right)^2dx\right)b+\left(\int_{x_1}^{x_2}\textrm{Win}\left(x\right)\tilde{E}_b\left(x\right)\tilde{E}_a\left(x\right)dx\right)a\right],\label{TCMTlinear_B}
\end{align}
\end{widetext}
which are just linear homogeneous equations of $\left\{a,b\right\}$.
The field integrals are constant independent of $\left\{a,b\right\}$ and $D_{\textrm{max}}$, so one can numerically calculate them with precomputed QNM field profiles $\tilde{E}_{a,b}\left(x\right)$.
Finding a non-zero solution to Eq.~\eqref{TCMTlinear_A}-\eqref{TCMTlinear_B} requires a vanishing determinant of the coefficient matrix.
Numerically solving this equation of vanishing determinant gives possible solutions of complex frequency $\omega$, which is plotted in Fig.~\ref{fig:stability}\textbf{b}.
Increasing pumping strength $D_{\textrm{max}}$ is equivalent to increase gain in this case.

Although the complex frequencies move toward the real axis as the pumping strength increases, they cannot go beyond the real axis as this would otherwise result in unphysical diverging field.
The critical point is when one of those complex frequencies first reaches the real axis, which marks the onset of the single-mode lasing.
The pumping strength at this point is labeled as the first lasing threshold $D_{\textrm{max}}=D_1^{th}$.

\section{\label{app:aboveD1}QNM expansion: single-mode lasing and stability analysis}
In the main text, the single-mode lasing is analyzed in a single cavity edge emitting laser, where only one QNM dominates the passive cavity response.
In this section, we consider the single-mode lasing in the coupled-cavity structure shown in Fig.~\ref{fig:4}, where both the two QNMs contribute to the single-mode lasing simultaneously.
The QNM expansion equations for single-mode lasing and its PALT stability analysis are derived.

%% alex recheck.

Beyond the first lasing threshold, the pump-induced gain balances the intrinsic loss of QNM and leads to stable single-mode lasing.
Since the loss of the QNM is solely determined by the passive resonator, the complete gain-loss cancellation requires the gain to remain constant while the pumping strength increases. 
This resistive mechanism in gain-pump relation is known as the gain saturation and can be expressed as a saturated polarization density
\begin{multline}
    \frac{d^2 E_0\left(x\right)}{dx^2}+\left(\frac{\omega_0}{c}\right)^2\varepsilon_r\left(x\right) E_0\left(x\right)= \\
    -\left(\frac{\omega_0}{c}\right)^2\frac{\Gamma_0 D_\textrm{P}\left(x\right)}{1+|\Gamma_0|^2|E_0\left(x\right)|^2}E_0\left(x\right),\label{eq:single}
\end{multline}
where $\Gamma_0=\Gamma_\perp\left(\omega_0\right)$.
Under the QNM expansion, the field is projected onto the 2-QNM basis as $E_0\left(x\right)=a_0\tilde{E}_a\left(x\right)+b_0\tilde{E}_b\left(x\right)$.
Observing that $P\left(x\right)$ is nonzero only inside the pumped cavity,
it is fair to assume that the two QNMs share the same field profile $E_\alpha\left(x\right)$ inside the pumped cavity 
\begin{equation}\label{eq:sameprofile}
    \tilde{E}_{a,b}\left(x\right)=\alpha_{1,2}E_\alpha\left(x\right), ~~\textrm{for }x\in \textrm{C}_{\textrm{act}},
\end{equation}
as plotted in Fig.~\ref{fig:4}\textbf{d} \& \textbf{e}. 
This assumption is guaranteed by the weak spatial coupling limit in such EP-laser \cite{gao2024dynamic}.
As a result, the QNM amplitude equations for the single-mode lasing operation can be simplified as
\begin{widetext}
    \begin{align}
    &\left(\tilde{\omega}_a^2-\omega_0^2\right)a_0\langle \tilde{E}_a|\tilde{E}_a \rangle=\omega_0^2\Gamma_0D_{\textrm{max}}\alpha_1\left(a_0\alpha_1+b_0\alpha_2\right)F\left(|\Gamma_0|^2|a_0\alpha_1+b_0\alpha_2|^2\right),\label{eq:TCMTsingle_A}\\
    &\left(\tilde{\omega}_b^2-\omega_0^2\right)b_0\langle \tilde{E}_b|\tilde{E}_b \rangle=\omega_0^2\Gamma_0D_{\textrm{max}}\alpha_2\left(a_0\alpha_1+b_0\alpha_2\right)F\left(|\Gamma_0|^2|a_0\alpha_1+b_0\alpha_2|^2\right).\label{eq:TCMTsingle_B}
    \end{align}
\end{widetext}

Due to the nonlinear saturation term in the denominator in ${P\left(x\right)}$, the integral is not a constant but a function of $\left\{a_0,b_0\right\}$,
\begin{equation}\label{eq:integralfunction}
    F\left(y\right)=\int_{\textrm{C}_{\textrm{act}}}\frac{\textrm{Win}\left(x\right)E_\alpha\left(x\right)^2}{1+|E_\alpha\left(x\right)|^2y} dx,
\end{equation}
where $y=|\Gamma_0|^2|a_0\alpha_1+b_0\alpha_2|^2$ and the integral interval $\textrm{C}_{\textrm{act}}$ is the pumped cavity.
Therefore, one only needs to precompute the QNMs $\tilde{E}_{a,b}\left(x\right)$ for $\left\{\alpha_1,\alpha_2,E_\alpha\left(x\right)\right\}$ in the integral function, which is just a standard nonlinear eigenvalue problem.
Consequently, Eq.~\eqref{eq:TCMTsingle_A}-\eqref{eq:TCMTsingle_B} are simply nonlinear equations of $\left\{a_0,b_0\right\}$, and thus can be solved using any basic nonlinear root-finding solver, as shown in Fig.~\ref{fig:stability}\textbf{h}.

The finite electric field of a single-mode lasing solution modifies the spatial distribution of population inversion as the gain saturates, yielding a change in the evolution of nonlasing modes as a function of pump strength.
The behavior of the lasing medium operating in the single-mode lasing regime can be understood using a linear stability analysis on the single-mode lasing solution \cite{gao2024dynamic}.
Here, we first perform the stability analysis on the single-mode solution, then simplify it using the QNM expansion.

%% alex here.

In presence of the single-mode lasing solution $E_0\left(x\right)e^{-i\omega_0 t}$ defined in Eq.~\eqref{eq:single}, a perturbation in the electric field with time dependence $e^{-i\left(\omega_0+\omega_d\right)t}$ will induce another signal with time dependence $e^{-i\left(\omega_0-\omega_d^*\right)t}$, where $\omega_d$ is the complex perturbation frequency.
Such a perturbed electric field $\delta E\left(x,t\right)=E_1\left(x\right)e^{-i\omega_1t}+E_{-1}\left(x\right)e^{-i\omega_{-1}t}$ is governed by a pair of coupled linear equations \cite{gao2024dynamic},
\begin{widetext}
\begin{align}
    \frac{d^2E_1\left(x\right)}{dx^2}+\left(\frac{\omega_1}{c}\right)^2\varepsilon_r E_1\left(x\right)&=-\left(\frac{\omega_1}{c}\right)^2\Gamma_{+}\left[(D_0\left(x\right)+\chi_+\left(x\right)E_0\left(x\right))E_1\left(x\right)+\chi_{-}\left(x\right)E_0\left(x\right)E_{-1}^*\left(x\right)\right],\label{eq:single_perturb1}\\
    \frac{d^2E_{-1}\left(x\right)}{dx^2}+\left(\frac{\omega_{-1}}{c}\right)^2\varepsilon_r E_{-1}\left(x\right)&=-\left(\frac{\omega_{-1}}{c}\right)^2\Gamma_{-}\left[(D_0\left(x\right)+\chi_{-}^*\left(x\right)E_0\left(x\right))E_{-1}\left(x\right)+\chi_{+}^*\left(x\right)E_0\left(x\right)E_{1}^*\left(x\right)\right],\label{eq:single_perturb2}
\end{align}
\end{widetext}
where $\omega_1=\omega_0+\omega_d$ and $\omega_{-1}=\omega_0-\omega_d^*$.
$\Gamma_{+}=\Gamma_\perp\left(\omega_1\right)$ and $\Gamma_{-}=\Gamma_\perp\left(\omega_{-1}\right)$ are the dispersive gain factors.
$D_0\left(x\right)=D_\textrm{P}\left(x\right)/\left[1+|\Gamma_\perp\left(\omega_0\right)|^2|E_0\left(x\right)|^2\right]$ is the saturated population inversion distribution at single-mode lasing operation. The additional two coupling coefficients are determined as
\begin{align}
    \chi_+\left(x\right)&=\frac{(1/2)\Gamma_{\parallel}\left(\Gamma_{+}-\Gamma_{0}^*\right)D_0\left(x\right)E_0^*\left(x\right)}{1-(1/2)\Gamma_{\parallel}\left(\Gamma_{+}-\Gamma_{-}^*\right)|E_0\left(x\right)|^2},\label{eq:chi1}\\
    \chi_{-}\left(x\right)&=\frac{(1/2)\Gamma_{\parallel}\left(\Gamma_{0}-\Gamma_{-}^*\right)D_0\left(x\right)E_0\left(x\right)}{1-(1/2)\Gamma_{\parallel}\left(\Gamma_{+}-\Gamma_{-}^*\right)|E_0\left(x\right)|^2}.\label{eq:chi-1}
\end{align}
Slightly beyond the first lasing threshold $D_1^{th}$, this nonlinear eigenvalue problem of complex $\omega_d$ has solutions with negative imaginary parts, which means the corresponding perturbation would decay exponentially.
Therefore, the single-mode lasing solution remains stable.

As the pumping strength $D_{\textrm{max}}$ increases, the solutions of $\omega_d$ moves on the complex plane until one of the solutions reaches the real axis.
A real eigenvalue of $\omega_d$ indicates a non-decaying perturbation and destabilizes the single-mode lasing solution.
This critical point of pumping strength is noted as the comb threshold $D_{\textrm{max}}=D_c^{th}$.

Under QNM expansion framework, the perturbation fields are also expanded onto the 2-QNM basis: $E_{\pm1}\left(x\right)=a_{\pm1}\tilde{E}_a\left(x\right)+b_{\pm1}\tilde{E}_b\left(x\right)$ and the corresponding nonlinear eigenvalue problem turns into
\begin{widetext}
    \begin{align}
    \left(\tilde{\omega}_a^2-\omega_1^2\right)a_1\langle \tilde{E}_a|\tilde{E}_a \rangle&=\omega_1^2\int_{\textrm{C}_{\textrm{act}}}\tilde{E}_a\left(x\right)\left\{\Gamma_{+}\left[\left(D_0\left(x\right)+\chi_+\left(x\right)E_0\left(x\right)\right)E_1\left(x\right)+\chi_{-}\left(x\right)E_0\left(x\right)E_{-1}^*\left(x\right)\right]\right\} dx,\label{TCMT_singleperturb1}\\
    \left(\tilde{\omega}_b^2-\omega_1^2\right)b_1\langle \tilde{E}_b|\tilde{E}_b \rangle&=\omega_1^2\int_{\textrm{C}_{\textrm{act}}}\tilde{E}_b\left(x\right)\left\{\Gamma_{+}\left[\left(D_0\left(x\right)+\chi_+E_0\left(x\right)\right)E_1\left(x\right)+\chi_{-}\left(x\right)E_0\left(x\right)E_{-1}^*\left(x\right)\right]\right\} dx,\label{TCMT_singleperturb2}\\
    \left(\tilde{\omega}_a^2-\omega_{-1}^2\right)a_{-1}\langle \tilde{E}_a|\tilde{E}_a \rangle&=\omega_{-1}^2\int_{\textrm{C}_{\textrm{act}}}\tilde{E}_a\left(x\right)\left\{\Gamma_{-}\left[\left(D_0\left(x\right)+\chi_{-}^*\left(x\right)E_0\left(x\right)\right)E_{-1}\left(x\right)+\chi_{+}^*\left(x\right)E_0\left(x\right)E_{1}^*\left(x\right)\right]\right\} dx,\label{TCMT_singleperturb3}\\
    \left(\tilde{\omega}_b^2-\omega_{-1}^2\right)b_{-1}\langle \tilde{E}_b|\tilde{E}_b \rangle&=\omega_{-1}^2\int_{\textrm{C}_{\textrm{act}}}\tilde{E}_b\left\{\Gamma_{-}\left[\left(D_0\left(x\right)+\chi_{-}^*\left(x\right)E_0\left(x\right)\right)E_{-1}\left(x\right)+\chi_{+}^*\left(x\right)E_0\left(x\right)E_{1}^*\left(x\right)\right]\right\} dx.\label{TCMT_singleperturb4}
    \end{align}
\end{widetext}
Despite the complicated nonlinear dependence of $E_0\left(x\right)$, Eq.~\eqref{TCMT_singleperturb1}-~\eqref{TCMT_singleperturb4} are intrinsically linear equations with respect to variables $\left\{a_1,b_1,a_{-1}^*,b_{-1}^*\right\}$.
Therefore, nonzero solution of $\left\{a_1,b_1,a_{-1}^*,b_{-1}^*\right\}$ requires vanishing determinant of the coefficient matrix, leading to complex eigenvalues $\omega_d$.

Noted that the integrals in Eq.~\eqref{TCMT_singleperturb1}-~\eqref{TCMT_singleperturb4} are also evaluated only inside $\textrm{C}_{\textrm{act}}$, the approximation of same field profile inside $\textrm{C}_{\textrm{act}}$ applies and the linear equations can be simplified as
\begin{widetext}
\begin{equation}\label{TCMT_singleperturb_matrix}
    \begin{bmatrix}  
      \omega_1^2\alpha_1I_1\alpha_1& \omega_1^2\alpha_1I_1\alpha_2& \omega_1^2\alpha_1I_2\alpha_1^*  & \omega_1^2\alpha_1I_2\alpha_2^* \\  
      \omega_1^2\alpha_2I_1\alpha_1& \omega_1^2\alpha_2I_1\alpha_2& \omega_1^2\alpha_2I_2\alpha_1^*  & \omega_1^2\alpha_2I_2\alpha_2^* \\  
      \left(\omega_{-1}^*\right)^2\alpha_1^*I_3^*\alpha_1& \left(\omega_{-1}^*\right)^2\alpha_1^*I_3^*\alpha_2& \left(\omega_{-1}^*\right)^2\alpha_1^*I_4^*\alpha_1^*  & \left(\omega_{-1}^*\right)^2\alpha_1^*I_4^*\alpha_2^* \\  
      \left(\omega_{-1}^*\right)^2\alpha_2^*I_3^*\alpha_1& \left(\omega_{-1}^*\right)^2\alpha_2^*I_3^*\alpha_2& \left(\omega_{-1}^*\right)^2\alpha_2^*I_4^*\alpha_1^*  & \left(\omega_{-1}^*\right)^2\alpha_2^*I_4^*\alpha_2^* 
    \end{bmatrix}  
    \begin{bmatrix}  
        a_1\\
        b_1\\
        a_{-1}^*\\
        b_{-1}^*
    \end{bmatrix}  
    -\textrm{diag}
    \begin{bmatrix}  
      \left(\tilde{\omega}_a^2-\omega_1^2\right)\langle \tilde{E}_a|\tilde{E}_a \rangle\\
      \left(\tilde{\omega}_b^2-\omega_1^2\right)\langle \tilde{E}_b|\tilde{E}_b \rangle\\
      \left(\tilde{\omega}_a^2-\omega_{-1}^2\right)^*\langle \tilde{E}_a|\tilde{E}_a \rangle^*\\
      \left(\tilde{\omega}_b^2-\omega_{-1}^2\right)^*\langle \tilde{E}_b|\tilde{E}_b \rangle^*
    \end{bmatrix}  
    \begin{bmatrix}  
        a_1\\
        b_1\\
        a_{-1}^*\\
        b_{-1}^*
    \end{bmatrix}=0,
\end{equation}
\end{widetext}
where $I_{1-4}$ consist of some integrals of the single-mode lasing solution $\left\{a_0,b_0,\omega_0\right\}$,
\begin{widetext}
\begin{align}
    \begin{split}
        I_1 = \Gamma_+ D_{\textrm{max}}\left\{F\left(|\Gamma_0|^2|a_0\alpha_1+b_0\alpha_2|^2\right)
        \left(
        -\frac{\Gamma_-^*-\Gamma_0^*}{\Gamma_+-\Gamma_-^*}+\frac{|\Gamma_0|^2}{|\Gamma_0|^2+(1/2)\Gamma_\parallel\left(\Gamma_+-\Gamma_-^*\right)}\frac{\Gamma_+-\Gamma_0^*}{\Gamma_+-\Gamma_-^*}
        \right)\right.\\
        \left.+F\left[-(1/2)\Gamma_\parallel\left(\Gamma_+-\Gamma_-^*\right)|a_0\alpha_1+b_0\alpha_2|^2\right]
        \frac{(1/2)\Gamma_\parallel\left(\Gamma_+-\Gamma_-^*\right)}{|\Gamma_0|^2+(1/2)\Gamma_\parallel\left(\Gamma_+-\Gamma_-^*\right)}\frac{\Gamma_+-\Gamma_0^*}{\Gamma_+-\Gamma_-^*}
        \right\},
    \end{split}\\
    \begin{split}
        I_2 = \Gamma_+ D_{\textrm{max}}\left\{F\left[-(1/2)\Gamma_\parallel\left(\Gamma_+-\Gamma_-^*\right)|a_0\alpha_1+b_0\alpha_2|^2\right]-F\left(|\Gamma_0|^2|a_0\alpha_1+b_0\alpha_2|^2\right)\right\}
        \frac{(1/2)\Gamma_\parallel\left(\Gamma_0-\Gamma_-^*\right)}{|\Gamma_0|^2+(1/2)\Gamma_\parallel\left(\Gamma_+-\Gamma_-^*\right)}
        \frac{a_0\alpha_1+b_0\alpha_2}{\left({a_0\alpha_1+b_0\alpha_2}\right)^*},
    \end{split}\\
    \begin{split}
        I_3 = \Gamma_- D_{\textrm{max}}\left\{F\left[-(1/2)\Gamma_\parallel^*\left(\Gamma_+^*-\Gamma_-\right)|a_0\alpha_1+b_0\alpha_2|^2\right]-F\left(|\Gamma_0|^2|a_0\alpha_1+b_0\alpha_2|^2\right)\right\}
        \frac{(1/2)\Gamma_\parallel^*\left(\Gamma_+^*-\Gamma_0\right)}{|\Gamma_0|^2+(1/2)\Gamma_\parallel^*\left(\Gamma_+^*-\Gamma_-\right)}
        \frac{a_0\alpha_1+b_0\alpha_2}{\left({a_0\alpha_1+b_0\alpha_2}\right)^*},
    \end{split}\\
    \begin{split}
        I_4 = \Gamma_- D_{\textrm{max}}\left\{F\left(|\Gamma_0|^2|a_0\alpha_1+b_0\alpha_2|^2\right)
        \left(
        -\frac{\Gamma_0^*-\Gamma_+^*}{\Gamma_+^*-\Gamma_-}+\frac{|\Gamma_0|^2}{|\Gamma_0|^2+(1/2)\Gamma_\parallel^*\left(\Gamma_+^*-\Gamma_-\right)}\frac{\Gamma_0^*-\Gamma_-}{\Gamma_+^*-\Gamma_-}
        \right)\right.\\
        \left.+F\left[-(1/2)\Gamma_\parallel^*\left(\Gamma_+^*-\Gamma_-\right)|a_0\alpha_1+b_0\alpha_2|^2\right]
        \frac{(1/2)\Gamma_\parallel^*\left(\Gamma_+^*-\Gamma_-\right)}{|\Gamma_0|^2+(1/2)\Gamma_\parallel^*\left(\Gamma_+^*-\Gamma_-\right)}\frac{\Gamma_0^*-\Gamma_-}{\Gamma_1^*-\Gamma_-}.
        \right\}
    \end{split}
\end{align}
\end{widetext}
Once the single-mode lasing solution $\{a_0,b_0,\omega_0\}$ is known, $I_{1-4}$ are just integral functions of $\omega_d$, and so is the determinant of the coefficient matrix in Eq.~\eqref{TCMT_singleperturb_matrix}.
Typical root-finding methods can be applied to find the zeros of this determinant, which corresponds to the complex perturbation frequency $\omega_d$.
By tracing $\omega_d$ while increasing $D_{\textrm{max}}$, the comb threshold $D_c^{th}$ can be found when $\omega_d$ reaches the real axis, marked as circles in Fig.~\ref{fig:stability}.

\section{\label{app:PALT}QNM-PALT: frequency comb solution for near EP lasers}
In this section, we apply the QNM expansion formalism to the EP comb generation and make an accurate prediction of the EP comb solution.

Above the comb lasing threshold $D_{\textrm{max}}>D_c^{th}$, the EP comb solution $E\left(x,t\right)=\sum_m E_m\left(x\right)e^{-i\omega_mt}$ are determined by the PALT equations \cite{gao2024dynamic},
\begin{widetext}
\begin{align}
    &\frac{d^2E_m\left(x\right)}{dx^2}+\frac{\omega_m^2}{c^2}\varepsilon_r\left(x\right)E_m\left(x\right)=-\frac{\omega_m^2}{c^2}\Gamma_{\perp}\left(\omega_m\right)\sum_{n=-\infty}^{+\infty}D_{m-n}\left(x\right)E_n\left(x\right),\label{SMeq:PALTE}\\
    &\bar{D}\left(x\right)=D_\textrm{p}\left(x\right)\left[\bar{\bar{I}}- (1/2) \bar{\bar{\Gamma}}_{\parallel}\left(\bar{\bar{E}}^\dag\left(x\right) \bar{\bar{\Gamma}}_+ \bar{\bar{E}}\left(x\right)-\bar{\bar{E}}\left(x\right) \bar{\bar{\Gamma}}_-^\dag \bar{\bar{E}}^\dag\left(x\right)\right)\right]^{-1}\bar{\delta},\label{SMeq:PALTD}
\end{align}
\end{widetext}
where $\omega_m=\omega_0+m\omega_d$ is the frequency of the mth Fourier component $E_m\left(x\right)$. 
The dynamic population inversion is expressed as $D\left(x,t\right)=\sum_n D_{n}\left(x\right)e^{-in\omega_dt}$. 
At equilibrium, the population inversion induced by electric field are determined by Eq.~\eqref{SMeq:PALTD}, where $\bar{D}\left(x\right)$ and $\bar{\delta}$ are column vectors with components $\left(\bar{D}\left(x\right)\right)_m=D_m\left(x\right)$ and $\left(\bar{\delta}\right)_m=\delta_{m0}$, $\delta$ is the Kronecker delta.
$\bar{\bar{E}}\left(x\right)$ is the electric field matrix filled with different Fourier components $\left(\bar{\bar{E}}\left(x\right)\right)_{mn}=E_{m-n}\left(x\right)$.
$\bar{\bar{\Gamma}}_{\parallel}$ and $\bar{\bar{\Gamma}}_{\pm}$ are diagonal matrices with $\left(\bar{\bar{\Gamma}}_{\parallel}\right)_{mn}=\delta_{m-n}\gamma_\parallel/\left(m\omega_d+i\gamma_\parallel\right)$ and $\left(\bar{\bar{\Gamma}}_{\pm}\right)_{mn}=\delta_{m-n}\Gamma_\perp\left(\omega_{\pm m}\right)$. 
$\bar{\bar{I}}$ is the identity matrix and $\dag$ stands for matrix conjugate transpose.

Expanding all of the Fourier components onto the 2-QNM basis, $E_m\left(x\right)=a_m\tilde{E}_a\left(x\right)+b_m\tilde{E}_b\left(x\right)$, the QNM amplitudes $\left\{a_m,b_m\right\}$ are determined by
\begin{widetext}
\begin{align}
    \left(\tilde{\omega}_a^2-\omega_m^2\right)a_m\langle \tilde{E}_a|\tilde{E}_a \rangle=\omega_m^2\int_{\textrm{C}_{\textrm{act}}}\tilde{E}_a\left(x\right)\Gamma_{\perp}\left(\omega_m\right)\sum_{n=-\infty}^{+\infty}D_{m-n}\left(x\right)E_n\left(x\right)dx,\label{TCMT_comba}\\
    \left(\tilde{\omega}_b^2-\omega_m^2\right)b_m\langle \tilde{E}_b|\tilde{E}_b \rangle=\omega_m^2\int_{\textrm{C}_{\textrm{act}}}\tilde{E}_b\left(x\right)\Gamma_{\perp}\left(\omega_m\right)\sum_{n=-\infty}^{+\infty}D_{m-n}\left(x\right)E_n\left(x\right)dx.\label{TCMT_combb}
\end{align}
\end{widetext}
For the integral inside $\textrm{C}_{\textrm{act}}$, we apply the same field profile approximation again and simplify the integrals as
\begin{widetext}
\begin{equation}
    \int_{\textrm{C}_{\textrm{act}}}\tilde{E}_{a,b}\left(x\right)\Gamma_{\perp}\left(\omega_m\right)\sum_{n=-\infty}^{+\infty}D_{m-n}\left(x\right)E_n\left(x\right)dx\approx
    D_{\textrm{max}}\alpha_{1,2}\sum_n M_{m-n}\left(a_n\alpha_1+b_n\alpha_2\right),
\end{equation}
\end{widetext}
where $M_{m-n}=\bar{M}_{m-n}$ and $\bar{M}$ is a column vector defined as
\begin{widetext}
\begin{equation}\label{M_matrix}
    \bar{M}=\bar{\bar{\Gamma}}_\perp\int_{\textrm{C}_{\textrm{act}}}\left[\bar{\bar{I}}+\bar{\bar{I}}_{\textrm{eff}}|E_\alpha\left(x\right)|^2\right]^{-1}\bar{\delta}\textrm{Win}\left(x\right)E_\alpha^2\left(x\right)dx,
\end{equation}
\end{widetext}
and $\bar{\bar{I}}_\textrm{eff}$ is the effective intensity matrix
\begin{widetext}
\begin{equation}
    \bar{\bar{I}}_{\textrm{eff}}=\frac{1}{2}\bar{\bar{\Gamma}}_\parallel\left[\left(\alpha_1\bar{\bar{A}}+\alpha_2\bar{\bar{B}}\right)\bar{\bar{\Gamma}}_-^\dag\left(\alpha_1^*\bar{\bar{A}}^\dag+\alpha_2^*\bar{\bar{B}}^\dag\right)-\left(\alpha_1^*\bar{\bar{A}}^\dag+\alpha_2^*\bar{\bar{B}}^\dag\right)\bar{\bar{\Gamma}}_+\left(\alpha_1\bar{\bar{A}}+\alpha_2\bar{\bar{B}}\right)\right],\label{I_matrix}
\end{equation}
\end{widetext}
and $\bar{\bar{A}}$, $\bar{\bar{B}}$ are the amplitude matrices $\left(\bar{\bar{A}}\right)_{mn}=a_{m-n}$ and $\left(\bar{\bar{B}}\right)_{mn}=b_{m-n}$.
The vectorial integral in Eq.~\eqref{M_matrix} is integrated for each element separately.

Equations~\eqref{TCMT_comba}-\eqref{TCMT_combb} are nonlinear integral equations of variables $\left\{a_m,b_m,\omega_0,\omega_d\right\}$ and are generally solvable with the well-developed root-finding techniques.
However, the presence of Eq.~\eqref{M_matrix} poses a great challenge in practical numerical computation due to repetitious evaluation of the matrix inverse at each position x and the subsequent integral for all matrix elements.

To simplify this numerical problem, we recognize that the effective intensity matrix Eq.~\eqref{I_matrix} is independent of space, thus can be diagonlized as space-independent eigenvalues and eigenvectors, $\bar{\bar{I}}_{\textrm{eff}}=\bar{\bar{P}}^{-1}\bar{\bar{\Lambda}}\bar{\bar{P}}$.
Substituting back into Eq.~\eqref{M_matrix}, we obtain
\begin{widetext}
\begin{equation}\label{M_matrix_diag}
    \begin{split}
    \bar{M}&=\bar{\bar{\Gamma}}_\perp\bar{\bar{P}}^{-1}\left\{\int_{\textrm{C}_{\textrm{act}}}\left[\bar{\bar{I}}+\bar{\bar{\Lambda}}|E_\alpha\left(x\right)|^2\right]^{-1}\textrm{Win}\left(x\right)E_\alpha^2\left(x\right)dx\right\}\bar{\bar{P}}\bar{\delta}\\
    &=\bar{\bar{\Gamma}}_\perp\bar{\bar{P}}^{-1}\bar{\bar{F}}\left(\bar{\bar{\Lambda}}\right)\bar{\bar{P}}\bar{\delta},
    \end{split} 
\end{equation}
\end{widetext}
where $\bar{\bar{F}}\left(\bar{\bar{\Lambda}}\right)$ is a matrix generalization of the scalar function $\left[\bar{\bar{F}}\left(\bar{\bar{\Lambda}}\right)\right]_{m,n}=F\left[\left(\bar{\bar{\Lambda}}\right)_{m,n}\right]$.
Compared with the original expression in Eq.~\eqref{M_matrix}, the space-independent matrix diagonalization in Eq.~\eqref{M_matrix_diag} is performed only once and the integral is computed only for the diagonal elements, while the space-dependent matrix inverse should be performed everywhere inside the integral interval $\textrm{C}_{\textrm{act}}$ and the integral is computed for every element in the matrix.

\section{\label{app:Pade}Pad\' e approximant}
Although the repetitive matrix inverse in Eq.~\eqref{M_matrix} is bypassed by a diagonalization approach in Eq.~\eqref{M_matrix_diag}, the evaluation of the integral function $F\left(y\right)$ in Eq.~\eqref{eq:integralfunction} can be the bottleneck of a computation.
In particular, the finite-difference simulation of this EP comb needs an abnormally small spatial grid size for convergence \cite{gao2024dynamic}, which makes the integral computation even slower.
Here we propose a numerical trick, which is a rational expansion of the integral function, to approximate the integral efficiently.

\begin{figure*}
    \centering
    \includegraphics[width=1\textwidth]{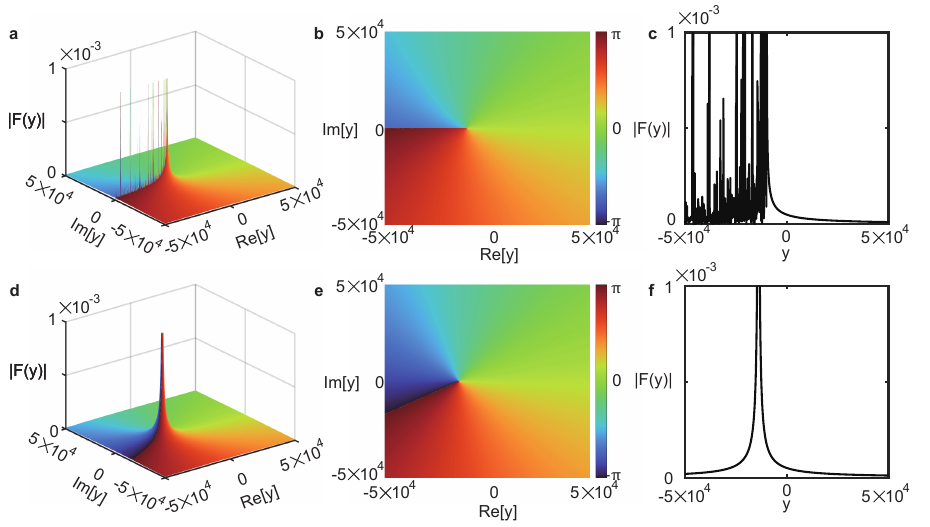}
    \caption{\textbf{a}, \textbf{d} Absolute value of $F(y)$ on the complex $y$ plane.
    \textbf{b}, \textbf{e} Phase of $F(y)$ on the complex $y$ plane.
    \textbf{c}, \textbf{f} Absolute value of $F(y)$ with real variable y.
    The first row are evaluated through the original integral function while the second row are evaluated as its Pad\' e approximant. 
    The uneven singular behavior in \textbf{a}-\textbf{c} is because of the finite sampling rate in the discretized electric field $E_\alpha(x_i)$.
    }
    \label{fig:Pade_comparison}
\end{figure*}

From the definition in Eq.\eqref{eq:integralfunction}, $F\left(y\right)$ is analytic on the entire complex y plane, except for a branch cut on the real axis $y=-1/|E_\alpha\left(x\right)|^2\textrm{, where } x\in \textrm{C}_{\textrm{act}}$.
As illustrated in Fig.~\ref{fig:Pade_comparison} \textbf{a}-\textbf{c}, $F(y)$ is a smooth function across the complex y plane with divergent behavior on part of the real axis.
In Fig.~\ref{fig:Pade_comparison} \textbf{a} \& \textbf{c} the non-analyticity seems discrete due to the discrete finite-difference data of $E_\alpha\left(x\right)$, and it should be continuous in general.

In the general context of nonlinear optics, the nonlinearity is always small and treated as perturbation \cite{boyd2008nonlinear}.
Therefore, a Taylor series expansion of the nonlinearity can be adopted, leading to the well known $\chi^{\left(2\right)}$, $\chi^{\left(3\right)}$, and higher order $\chi^{\left(n\right)}$ effects.
However, the nonlinearity in laser medium originates mainly from the mode competition caused by gain saturation, which could deprive most of the gain and leave very little gain for the rest modes when operating at high intensities.
Such vanishing behavior of $F\left(y\right)$ at infinity intensity falls outside the radius of convergence of Taylor expansion thus cannot be captured by a Taylor series expansion in general.

Instead of a Taylor series expansion, here we use a rational function expansion
\begin{equation}
    F\left(y\right)=\int_{\textrm{C}_{\textrm{act}}}\frac{\textrm{Win}\left(x\right)E_\alpha\left(x\right)^2}{1+|E_\alpha\left(x\right)|^2y} dx\approx\frac{P\left(y\right)}{Q\left(y\right)}
\end{equation}
to estimate the function value.
$P\left(y\right)$, $Q\left(y\right)$ are polynomial function of $y$.
Such estimation is also referred to as the Pad\' e approximant of $F\left(y\right)$ \cite{baker1961pade,baker1964theory}.
We find that the the simplest form of $F\left(y\right)\approx\frac{\lambda}{1+\mu y}$ is sufficient to provide a rather good estimate of $F(y)$.
$\left\{\lambda,\mu\right\}$ are complex constants and can be fitted from some precomputation.
For instance, one can evaluate $F\left(y\right)$ for real valued $y$ then fit the result with $F\left(y\right)\approx\frac{\lambda}{1+\mu y}$, as illustrated in Fig.~3\textbf{a}-\textbf{b} in the main text.

\begin{figure}
    \centering
    \includegraphics[width=0.5\textwidth]{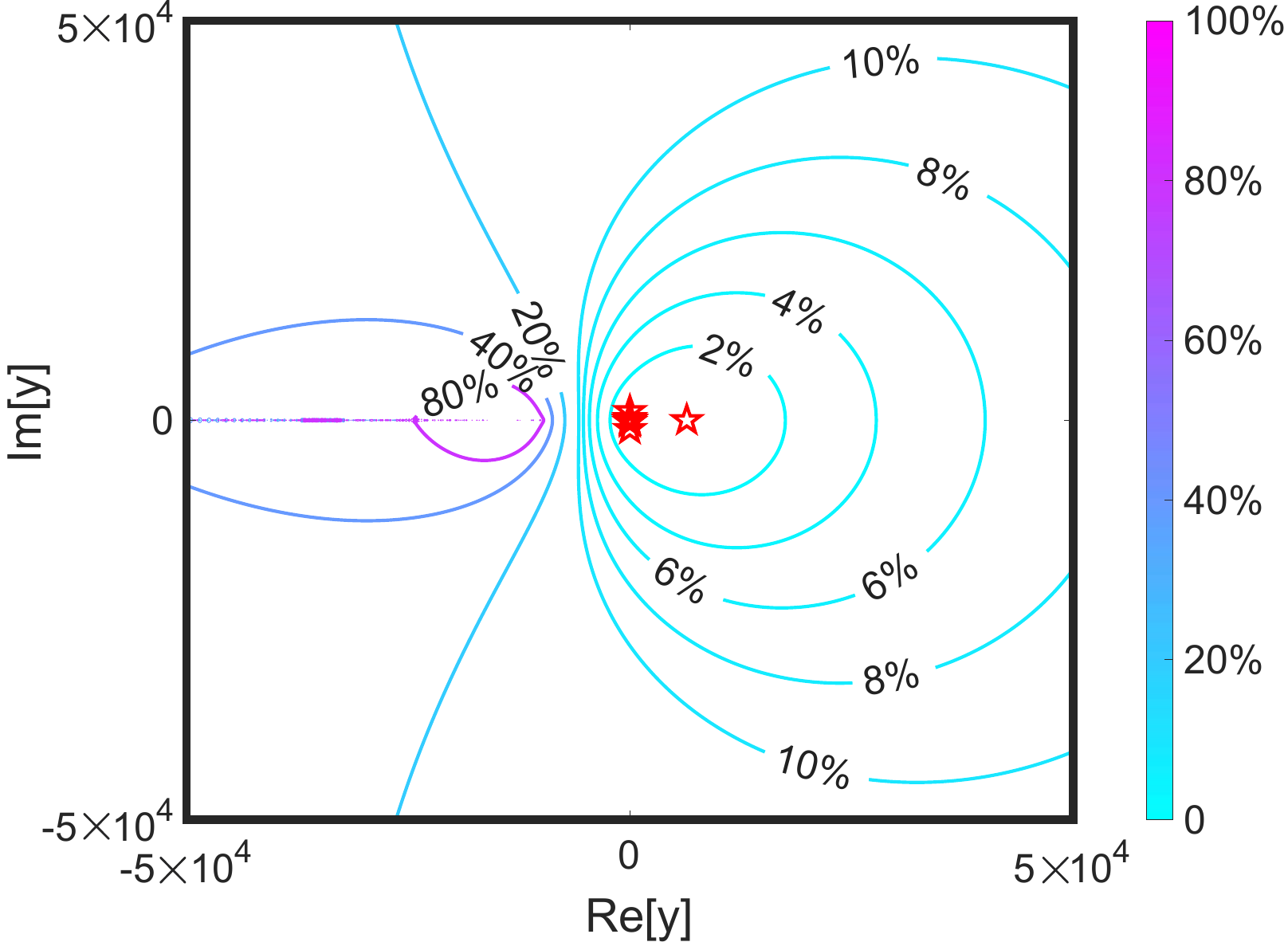}
    \caption{The error distribution of a Pad\' e approximant in $F(y)$ evaluation on the complex $y$ plane.
    The error is defined as the relative difference between the approximant and the exact value: $\textrm{error}=|\left(F_{\textrm{approximant}}-F_{\textrm{original}}\right)/F_{\textrm{original}}|$.
    The red pentagrams marks the variables of $F\left(y\right)$ in the comb solution shown in Fig.~\ref{fig:5} \textbf{f}.}
    \label{fig:padeerror}
\end{figure}
As shown in Fig.~\ref{fig:Pade_comparison}\textbf{d}-\textbf{f}, the new rational function $F\left(y\right)\approx\frac{\lambda}{1+\mu y}$ only has one single pole $y=-1/\mu$ around the negative real axis, which serves as a weighted average over all the poles of original $F\left(y\right)$.
The error distribution of this estimation is plotted as contour lines on the complex $y$ plane in Fig.~\ref{fig:padeerror}.
An accuracy of $10\%$ is achieved for most regions with a positive real part.
Moreover, the practical complex variable y in laser operations resides near the positive real axis, which leads to much higher accuracy.
For instance, in the single-mode lasing operation, $y=|\Gamma_0|^2|a_0\alpha_1+b_0\alpha_2|^2$ is real, this yielding high accuracy.
For the EP comb solution, the values of $y$, which are the eigenvalues of the effective intensity matrix $\bar{\bar{I}}_{\textrm{eff}}$, are plotted as red pentagrams in {Fig.}~\ref{fig:padeerror}.
All of these eigenvalues are enclosed by the contour line of $2\%$ relative error.
Such a low level of error ensures a good estimation of $F\left(y\right)$, leading to an accurate and efficient way to evaluate the integral function and perform PALT-related computations.
% \bibliographystyle{naturemag}
% \bibliography{achemso-demo}

\bibliography{apssamp}% Produces the bibliography via BibTeX.

\end{document}